\documentclass[review,1p
]{elsarticle}

\usepackage[utf8]{inputenc}
\usepackage[T1]{fontenc}
\usepackage{amsmath}
\usepackage{amssymb}
\usepackage{eucal}
\usepackage{graphicx}
\usepackage{aas_macros}
\usepackage[tight]{subfigure}
\usepackage{multirow}
\usepackage{booktabs}

\newcommand{\un}[1]{\,{\rm #1}}
\newcommand{\dd}{\mathrm{d}}
\newcommand{\degrees}{^\circ}
\newcommand{\eps}{\varepsilon}
\DeclareMathOperator{\cov}{cov}

\journal{Astroparticle Physics}

\begin{document}

\begin{frontmatter}

\title{All-particle cosmic ray energy spectrum measured with 26 IceTop stations}

\author[Madison]{R.~Abbasi}
\author[Gent]{Y.~Abdou}
\author[RiverFalls]{T.~Abu-Zayyad}
\author[Zeuthen]{M.~Ackermann}
\author[Christchurch]{J.~Adams}
\author[Geneva]{J.~A.~Aguilar}
\author[Madison]{M.~Ahlers}
\author[Aachen]{D.~Altmann}
\author[Madison]{K.~Andeen}
\author[Madison]{J.~Auffenberg}
% ...
\address[Aachen]{III. Physikalisches Institut, RWTH Aachen University, D-52056 Aachen, Germany}
% ...
\address[Christchurch]{Dept.~of Physics and Astronomy, University of Canterbury, Private Bag 4800, Christchurch, New Zealand}
% ...
\address[Geneva]{D\'epartement de physique nucl\'eaire et corpusculaire, Universit\'e de Gen\`eve, CH-1211 Gen\`eve, Switzerland}
\address[Gent]{Dept.~of Physics and Astronomy, University of Gent, B-9000 Gent, Belgium}
% ...
\address[Madison]{Dept.~of Physics, University of Wisconsin, Madison, WI 53706, USA}
% ...
\address[RiverFalls]{Dept.~of Physics, University of Wisconsin, River Falls, WI 54022, USA}
% ...
\address[Zeuthen]{DESY, D-15735 Zeuthen, Germany}
\fntext[SouthDakota]{Physics Department, South Dakota School of Mines and Technology, Rapid City, SD 57701, USA}
\fntext[LosAlamos]{Los Alamos National Laboratory, Los Alamos, NM 87545, USA}
\fntext[Bari]{also Sezione INFN, Dipartimento di Fisica, I-70126, Bari, Italy}
\fntext[Goddard]{NASA Goddard Space Flight Center, Greenbelt, MD 20771, USA}

\cortext[cor]{Corresponding author}

\begin{abstract}
We report on a measurement of the cosmic ray energy spectrum with the IceTop air shower array, the surface component of the IceCube Neutrino Observatory at the South Pole. 
The data used in this analysis were taken between June and October,~2007, with~26 surface stations operational at that time, corresponding to about one third of the final array.
The fiducial area used in this analysis was~$0.122\un{km^2}$.
The analysis investigated the energy spectrum from~$1$ to~$100\un{PeV}$ measured for three different zenith angle ranges between~$0\degrees$ and~$46\degrees$. 
Because of the isotropy of cosmic rays in this energy range the spectra from all zenith angle intervals have to agree.
The cosmic-ray energy spectrum was determined under different assumptions on the primary mass composition.
Good agreement of spectra in the three zenith angle ranges was found for the assumption of pure proton and a simple two-component model.
For zenith angles~\mbox{$\theta < 30\degrees$}, where the mass dependence is smallest, the knee in the cosmic ray energy spectrum was observed between~$3.5$ and~$4.32\un{PeV}$, depending on composition assumption.
Spectral indices above the knee range from~$-3.08$ to~$-3.11$ depending on primary mass composition assumption.
Moreover, an indication of a flattening of the spectrum above~$22\un{PeV}$ were observed.
\end{abstract}



\end{frontmatter}

\section{Introduction}
Almost~100 years after the discovery of cosmic rays, their sources and acceleration mechanisms still remain mostly unknown.
The energy spectrum of cosmic rays as measured by various experiments follows a relatively smooth power law with spectral index~$\gamma \approx -2.7$ up to about $4\un{PeV}$, where it steepens to \mbox{$\gamma \approx -3.1$}~\cite{Hoerandel_on_the_knee}.
While this feature in the spectrum called ``knee'' is well established, its origin remains controversial~\cite{Hoerandel_models_of_the_knee}.
Most models to explain the knee involve a change in chemical composition of cosmic rays in the energy region above the knee.
Such a change has been observed by various experiments~\cite{pdg:2010} but systematic uncertainties are too large to discriminate individual descriptions.
Features in the all-particle cosmic ray energy spectrum and their chemical composition bear important information on the acceleration and propagation of cosmic rays.
The measurement of the cosmic ray energy spectrum and composition is the main goal of the IceTop air shower array.

IceTop is the surface component of the IceCube Neutrino Observatory at the geographic South Pole. 
Installation of IceCube and IceTop was completed at the end of 2010, with 86 IceCube strings and 81 IceTop stations deployed covering an area of about~$1\un{km^2}$ and a volume of about~$1\un{km^3}$. 
IceTop was designed to measure the energy spectrum and the primary mass composition of cosmic ray air showers in the energy range between $5\cdot 10^{14}\un{eV}$ and $10^{18}\un{eV}$. 

The average atmospheric depth at the South Pole is about $680\un{g/cm^2}$. 
IceTop is therefore located close to the shower maximum for showers in the PeV range (for vertical protons about $550\un{g/cm^2}$ at $1\un{PeV}$ to $720\un{g/cm^2}$  at $1\un{EeV}$). 
This has the advantage that local shower density fluctuations are smaller than at later stages of shower development.

In this paper, we present the first analysis of IceTop data on high-energy cosmic rays and a measurement of the cosmic ray energy spectrum. 
This analysis is based on air shower data taken with the IceTop surface stations. 
The data were taken between June and October 2007 with 26 IceTop stations operating, which comprise about 1/3 of the complete detector.

Section~\ref{sec:detector} of this paper gives an overview over the IceTop array, and the processing and calibration of tank signals, which are the basis for reconstructing air showers. 
Section~\ref{sec:dataset} describes the dataset and run selection criteria.
Section~\ref{sec:reconstruction} introduces event reconstruction, and in Section~\ref{sec:simulation}, simulation of air showers and of the IceTop tank response are presented. 
In Section~\ref{sub:final_selection} the final event selection and detector performance are discussed.
Section~\ref{sec:energy_reco} describes the determination of the primary energy, whereas systematic uncertainties are discussed in Section~\ref{sec:systematics}.
In Section~\ref{sec:result} the results are presented and discussed.

\section{The detector}\label{sec:detector}

\subsection{IceTop}

\begin{figure*}
  \centering
  \includegraphics[width=.7\textwidth]{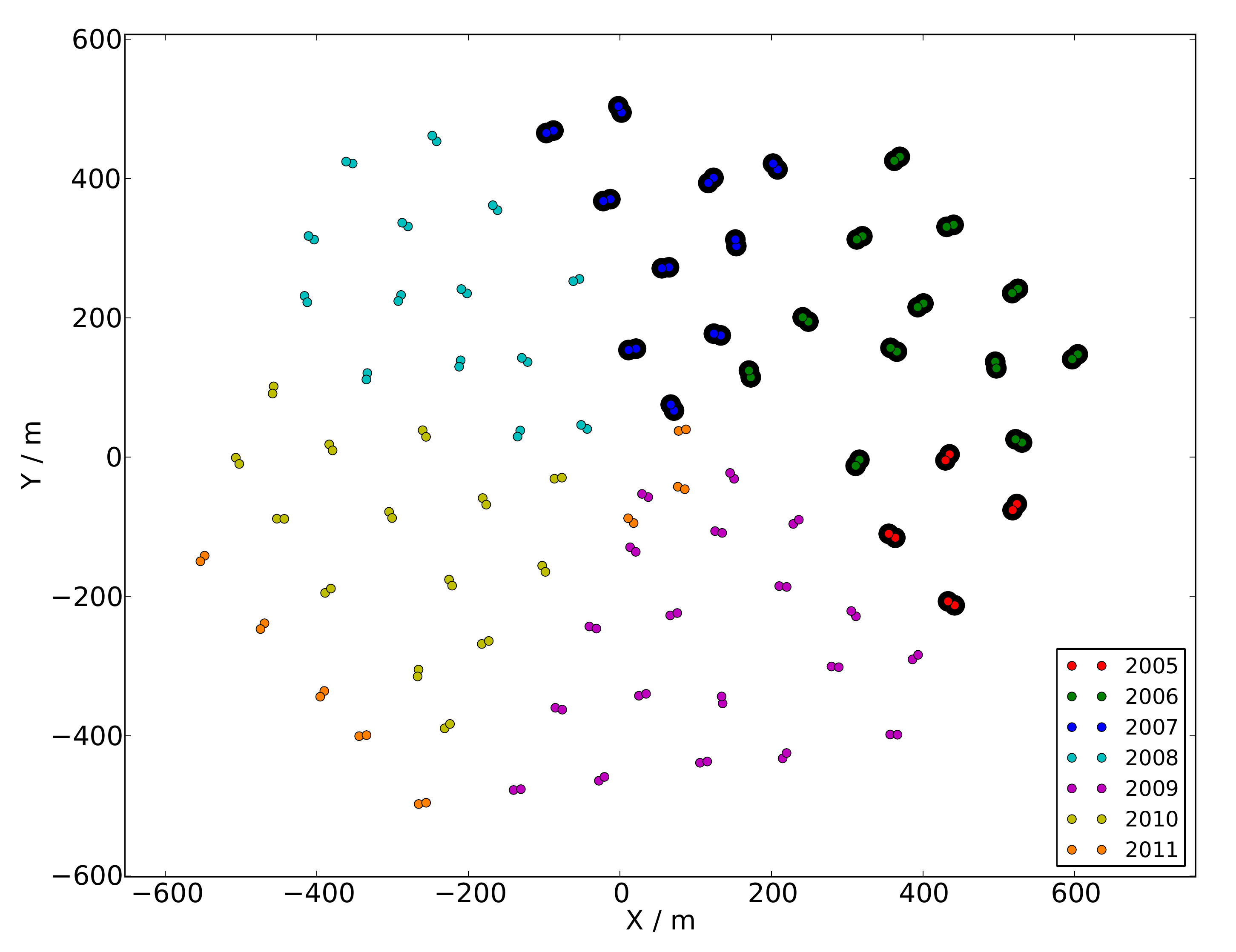}
  \caption{Layout of the IceTop air shower array. Colors indicate the year of deployment and the 26 stations installed in 2007 are highlighted.}
  \label{fig:array}
\end{figure*} 

The IceTop air shower array is the surface component of IceCube, covering an area of about $1\un{km^2}$ with 81 detector stations above the 86 IceCube strings.
The stations are mostly located next to IceCube strings with a average spacing of~$125\un{m}$, except for three stations placed as an infill with a smaller spacing in the central part of the detector, in order to lower the energy threshold of the detector to about~$100\un{TeV}$.
By~2007,~22 IceCube strings and~26 IceTop stations had been deployed.
These stations are highlighted in Fig.~\ref{fig:array}, which shows the layout of the IceTop air shower array in its final configuration.

Each station consists of two ice-filled tanks separated from each other by~$10\un{m}$. 
The two tanks of each station are embedded in snow with their tops aligned with the surface in order to minimize the accumulation of drifting snow (see Section~\ref{sub:snow}) and to protect the ice from temperature variations.

\begin{figure}
  \centering
  \includegraphics[width=1.00\columnwidth]{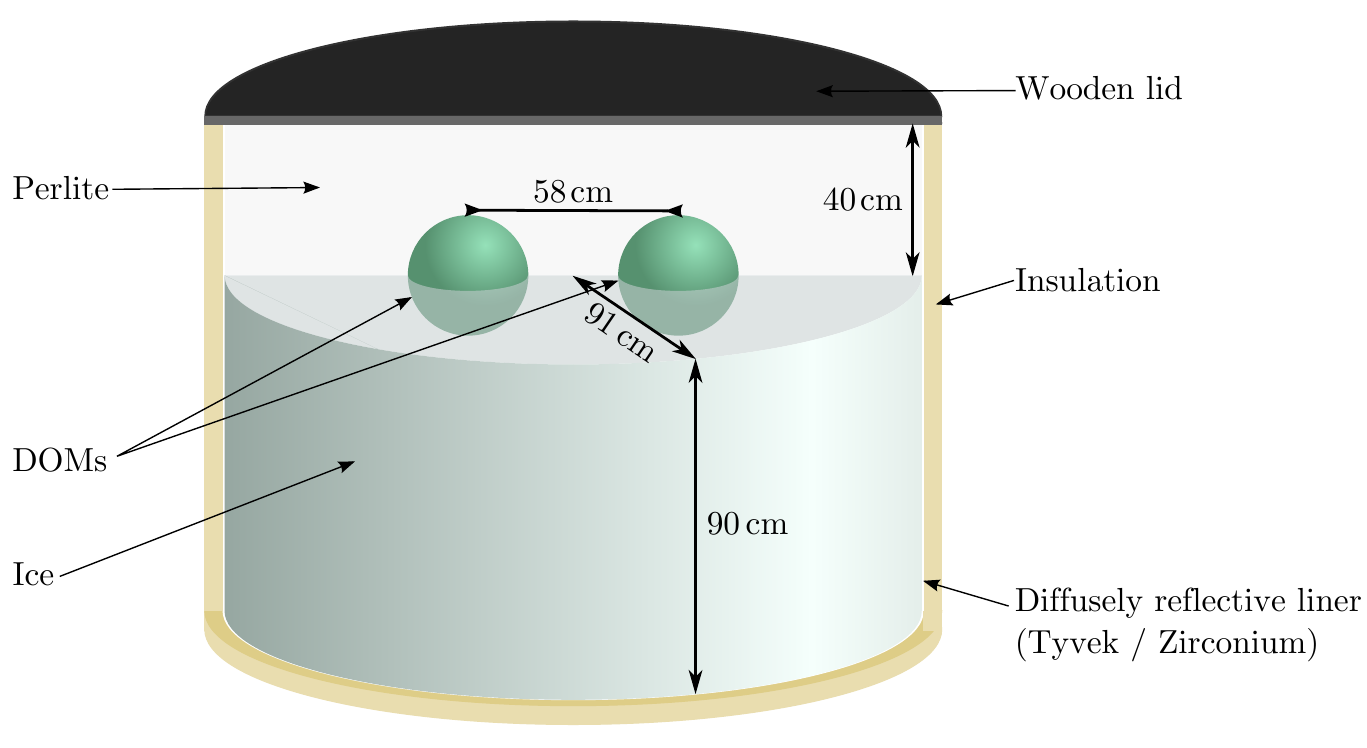}
  \caption{Cross section of a tank showing the tank geometry with insulation and position of the DOMs. The center of the ice surface between the two DOMs is used as tank position by reconstruction algorithms.}
  \label{fig:tank}
\end{figure}

The tanks are cylindrical with an inner diameter of~$1.82\un{m}$, and are filled with transparent ice to a depth of~$90\un{cm}$ (see Fig.~\ref{fig:tank}). 
The inner tank walls are covered with a diffusely reflective zirconium coating.
The first four stations deployed in 2005 and four tanks of the infill have a Tyvek liner with a higher reflectivity.
This difference affects amplitude and pulse width of detected tank signals, since the higher reflectivity reduces Cherenkov photon absorption, leading to longer pulses.

Each tank is equipped with two `Digital Optical Modules' (DOMs)~\cite{DOMPaper} to record Cherenkov light generated by charged particles passing through the tank.
The DOMs are identical to those used in other IceCube components and consist of a~$10''$ photomultiplier tube (PMT)~\cite{PMTPaper}, plus electronic circuitry for signal digitization, readout, triggering, calibration, data transfer and various control functions. 
The two DOMs in each tank were operated at different PMT gains, $5 \cdot 10^6$ (high-gain DOM) and $5 \cdot 10^5$ (low-gain DOM), to enhance the dynamic range.
This resulted in a linear dynamic range from $1$ to more than $10^5$ photoelectrons (PE).
During the data taking period used in this analysis all 104 DOMs in the 26 IceTop stations were fully operational.

\subsection{Trigger and data acquisition}\label{sub:trigger}
A DOM records PMT signals autonomously. 
A signal is recorded if it surpasses a certain discriminator threshold, which in the case of IceTop was set to~$22\un{mV}$ for the high-gain DOMs (corresponding to about~$20\un{pe}$) and~$12\un{mV}$ for the low-gain DOMs (corresponding to about~$180\un{pe}$).
The exact charge threshold depends on the pulse shape, which is determined by the arrival times of photoelectrons. 
After triggering, the delayed PMT pulse is sampled by `Analog Transient Waveform Digitizers' (ATWDs) with three different gain channels (nominal gains are 0.25, 2, and 16) in~128 bins with a width of~$3.3\un{ns}$, corresponding to a total sampling time of about~$422\un{ns}$. 
The analog samples are then digitized to~10~bits accuracy.

Up to this point, signal recording happens independently in each DOM. 
To reduce the high trigger rates in high-gain DOMs~(${\sim}2\un{kHz}$), which are mostly from low-energy showers, a hardware `local coincidence' between the high-gain DOMs in the two tanks of a station is required to initiate the readout and transmission of DOM data to the counting house (IceCube Lab). 
The digitizing process is aborted if the high-gain DOM in the neighboring tank does not also measure a signal above threshold within a time window of~$\pm 1\un{\mu s}$. 
The IceTop trigger condition is satisfied, if six or more DOMs report a (local coincident) signal within a time window of~$5\un{\mu s}$, which initiates readout of all DOMs from~$10\un{\mu s}$ before the first until $10\un{\mu s}$ after the last of the six DOM triggers which initiated the readout.
The requirement of 6 DOMs means that at least two stations had to trigger.
In 2007, the total IceTop trigger rate was about~$14\un{Hz}$.

\subsection{Charge extraction and calibration} \label{sub:calibration}
\begin{figure*}%
  \centering%
  \includegraphics[width=\textwidth]{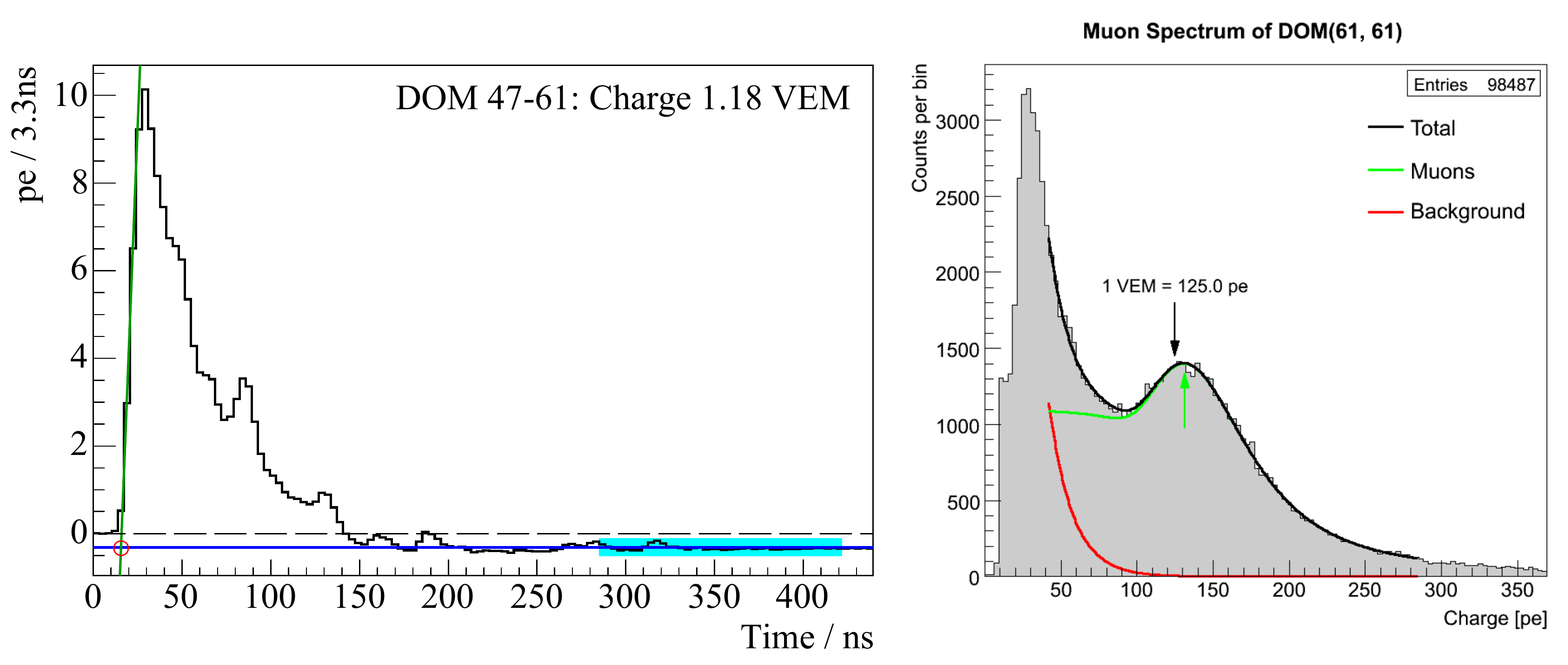}%
  \caption{\emph{Left:} A typical IceTop waveform. The blue horizontal line marks the baseline and the near vertical green line indicates the extrapolation of the leading edge yielding the signal time marked by the red circle. The baseline is below 0 (dashed line) due to droop. \emph{Right:} A typical charge spectrum recoded for the VEM calibration. The spectra are fitted with an empirical formula to determine the peak position (see text).}
  \label{fig:waveform-vemcal}
\end{figure*}

Figure~\ref{fig:waveform-vemcal} shows a typical waveform measured in IceTop.
While waveforms are recorded in three ATWD channels, this analysis used only the highest gain unsaturated (less than~1022~ADC counts) channel.
In this analysis only the integrated charge and the signal time were used. 
Before a waveform was integrated, its baseline was subtracted by determining the average value in bins 83 to 123 highlighted in the figure.
The undershoot is caused by droop introduced by the ferrite-core transformer used to couple the photomultiplier tube to the DOM's front-end electronics.
The signal time (`leading edge time') was defined by extrapolating the steepest rise of the waveform before the maximum down to the baseline. 
The absolute time scale of a DOM is calibrated with respect to all other DOMs to an accuracy of about~$2\un{ns}$~RMS~\cite{first_year_icecube}.

The charge produced by a single photoelectron, the amplifier gains and the digitizers are calibrated in a procedure common to all IceCube DOMs~\cite{first_year_icecube}.
However, the signal response to a particle of a given type and energy traversing the tank, expressed in photoelectrons, differs from tank to tank, due to differences in ice quality and reflectivity of the tank walls. 
Therefore, the signal of each tank is converted to a common unit called `Vertical Equivalent Muon' (VEM). 
Calibration was done by recording charge spectra of DOMs in dedicated calibration runs with all DOMs operated at a gain of $5 \cdot 10^6$ and without requiring local coincidence (for an example see Fig.~\ref{fig:waveform-vemcal}, right). 
These charge spectra show a clear peak due to penetrating muons above a background of electrons and photons. 
The spectra are fitted by the sum of a function describing the muon peak and an exponentially falling background term.
Measurements with a portable scintillator telescope mounted on top of tanks, restricting muons to nearly vertical angles of incidence, indicated that the muon peak lies about~$5\%$ lower than for the full angular range.
Simulation studies confirmed that restricting the angles of incidence of muons shifts the peak position by about~$5\%$~\cite{demiroers07}.
The scaled peak is referred to as `VEM peak'. 
For a given DOM the VEM unit can be expressed in terms of number of photoelectrons.
These values average~$120$ and~$200$ photoelectrons for the low and high reflectivity tanks (see above), respectively. 

For the 5-month run, 15 calibration runs were used. 
Between two consecutive calibration runs, the charge calibration was assumed to be stable (see also the discussion in Section~\ref{sub:systematics-vem}).

\subsection{Atmospheric conditions} \label{sub:atmosphere}
Variations of the atmosphere influence the development of air showers and thus the signals measured in IceTop.
Since IceTop is below the shower maximum for all energies of interest in this analysis and for all primary masses, an increase of the atmospheric overburden leads to an attenuation of shower sizes.
Atmospheric overburden is related to ground pressure~$p$ as~\mbox{$X_0 = p/g$}, where~\mbox{$g = 9.87\un{m/s^2}$} is the gravitational acceleration at the South Pole.
While there is some annual variation of the ground pressure, it mostly varies on shorter time scales on the order of days.

Besides ground pressure, the altitude profile of the atmosphere, $\dd X_v(h)/\dd h$, also influences the development of air showers.
This altitude profile has a pronounced annual cycle because the cold atmosphere during the winter months is much denser than the warmer atmosphere of the summer months.
The data used in this analysis were mostly taken during the winter months.

In the simulations used to interpret the air shower data a model of the South Pole atmosphere is used, which should represent the average atmosphere during the data taking period.
Nevertheless, variations of the atmosphere around the average lead to an additional uncertainty on the measured energy spectrum.
These systematic uncertainties will be discussed in Section~\ref{sub:systematics:atm_variation}.

\subsection{Snow} \label{sub:snow}
During installation, IceTop tanks are embedded in snow up to the upper surface of the tanks. 
Depending on location, surrounding surface and structures, each tank is covered by accumulated layers of snow of varying thickness.
Each year the amount of snow on the IceTop tanks grows by an average of $20\un{cm}$.

\begin{figure}
  \centering
  \includegraphics[width=.6\columnwidth]{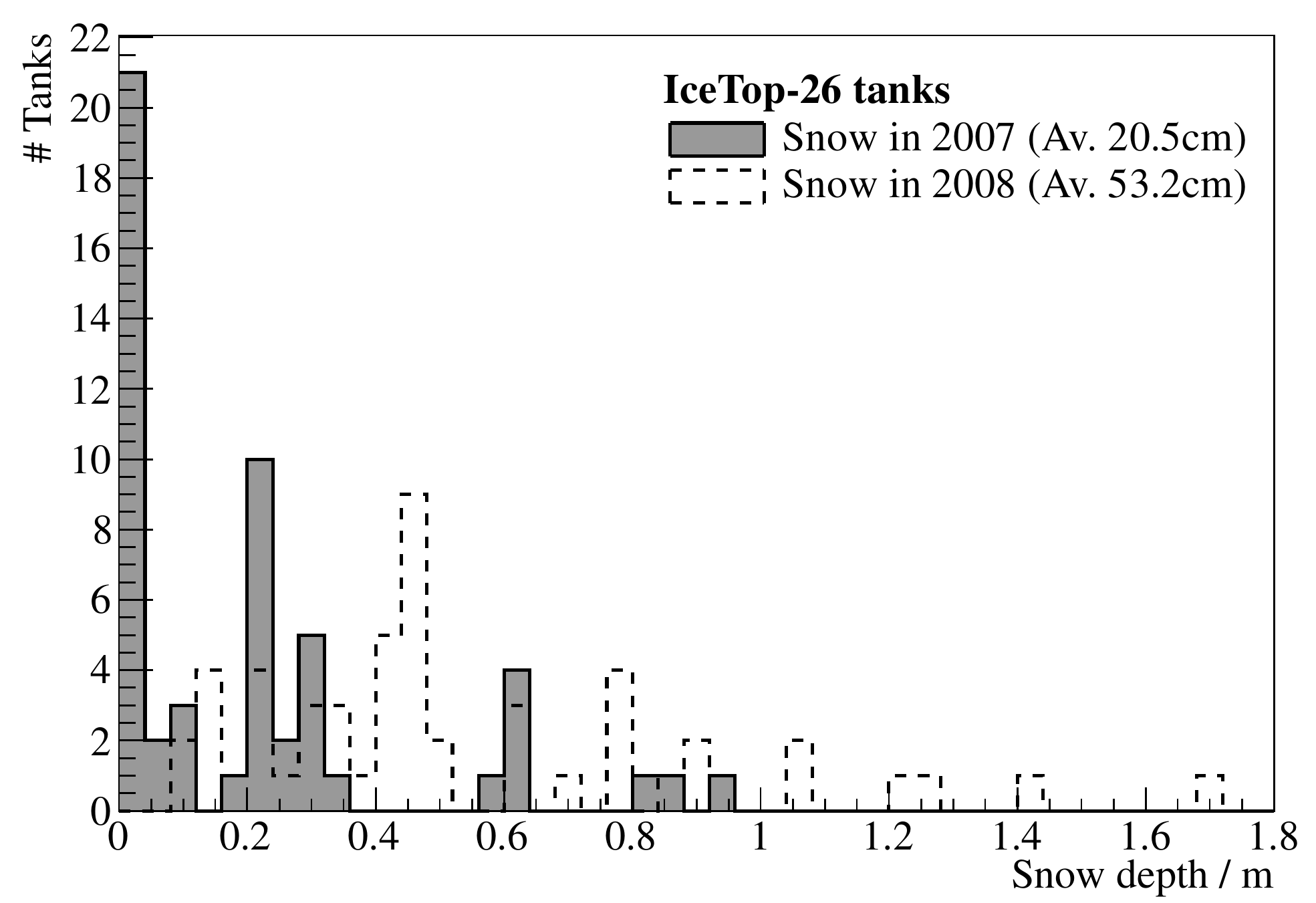}
  \caption{Snow heights on top of IceTop tanks measured in January~2007. All 20 newly deployed tanks had no snow on top, the average snow height was $20.5\un{cm}$. The dashed histogram is the snow height distribution on top of the same tanks measured one year later.}
  \label{fig:snowheights}
\end{figure}

As shown in Fig.~\ref{fig:snowheights}, the snow height for the analyzed data varied mostly between~$0$ and~$30\un{cm}$, except for four stations close to a building, which are covered by~$60$ to~$90\un{cm}$ of snow. 
The average snow height was~$20.5\un{cm}$ in January,~2007.

The snow has an average density of $0.38\un{g/cm^3}$, depending on snow height and location. 
The snow on top of and around the tanks influences the response to air shower particles penetrating the tanks and needs to be taken into account in simulations and for the determination of the shower energy.

\section{Data set and data selection}\label{sec:dataset}

\paragraph{Event filtering and data transmission}
The data used in this analysis were taken between June 1st and October 31st, 2007.
The analysis was performed using a data sample which was transferred via satellite to the IceCube data center at UW Madison with limited bandwith.
Due to these bandwidth constraints, events with less than~16 participating DOMs were prescaled by a factor of~5.
Events with 16 or more DOMs were transmitted at a rate of $0.9\un{Hz}$ and the small events at a rate of $2.5\un{Hz}$.

\paragraph{Run selection}
In order to ensure detector stability and data quality, the following criteria were applied to runs which were used in this analysis:
\begin{itemize}
 \item The run was longer than $30\un{min}$. A normal detector run lasted $8$ hours, and nearly all runs that were aborted after a short time encountered some sort of problem.
 \item All DOMs were running stably.
 \item After correction for atmospheric pressure variations the trigger and filter rates were stable and within $\pm 5\%$ agreement with the previous good run. Pressure correction was done by fitting the relation between ground pressure $p$ and rate $R$ with an exponential function, $R(p) \sim \exp(-\beta \, p)$, yielding a barometric coefficient $\beta = 0.0077/\mathrm{mbar}$~\cite{Tilav09}. Then, the rates were corrected to the average South Pole ground pressure of $680\un{mbar}$:
 \begin{equation}
   R_\mathrm{corrected} = R \, \exp\bigl(\beta \, (p-680\un{mbar}) \bigr).
 \end{equation}
\end{itemize}
These cuts reduced the livetime by about $10\%$.

\paragraph{Event cleaning}
Before starting the reconstruction, events were cleaned based on a few simple timing criteria.
In case both DOMs of a tank triggered, the tank signal was rejected if the time difference between the two signals was greater than $40\un{ns}$.
The analysis used only one signal per tank.
For each high-gain DOM a saturation threshold was determined from a comparison of signals that triggered both DOMs in a tank.
Signals with less charge were taken from the high-gain DOM.
If the charge exceeded the saturation threshold, the charge measured by the low-gain DOM was used and the time was determined from the high-gain signal.
Furthermore, a tank signal was also rejected if only the low-gain DOM triggered and the high-gain DOM was missing.

Then, a maximum time difference of
\begin{equation}
  |t_A - t_B| < \frac{|\boldsymbol{x_A}-\boldsymbol{x_B}|}{c} + 200\un{ns}
\end{equation}
between signals in tanks A and B of the same station was required.
Here, $t_A$ and $t_B$ are the signal times in the two tanks and $\boldsymbol{x_A}$ and $\boldsymbol{x_B}$ are the tank locations.
The tolerance of $200\un{ns}$ was introduced in order to account for shower fluctuations.
Finally, stations were grouped in clusters, such that any pair of stations $i$ and $j$ in the cluster fulfilled the condition
\begin{equation}
  |t_i - t_j| < \frac{|\boldsymbol{x_i}-\boldsymbol{x_j}|}{c} + 200\un{ns}.
\end{equation}
The station position $\boldsymbol{x_i}$ is the center of the line connecting its two tanks, and $t_i$ is the average time of the tank signals.
In each event, only the largest cluster of stations was kept.

Only about $10\%$ of events were affected by this event cleaning, and about~$2.5\%$ of events dropped below the threshold of 5 stations required for reconstruction.

\paragraph{Charge-based retriggering}\label{sub:retriggering}
In order to reduce uncertainties due to the description of the detector threshold in the simulation, all events were retriggered to a common threshold based on total registered charge. 
All pulses with a charge below $S_{\rm thr} = 0.3\un{VEM}$ were removed, and afterwards the local coincidence conditions (see Section~\ref{sub:trigger}) were re-evaluated discarding all pulses that no longer fulfilled this condition. 
This procedure was applied to both experimental and simulated data.

\paragraph{Event selection}
For further processing, a total of $N_\mathrm{tot} = 8\,895\,205$ events were selected where at least five stations had triggered. 
Events which fulfilled this condition, but had less than 16 DOMs read out (before event cleaning), were reweighted in the analysis with the prescale factor of~5~(see above). 

The effective livetime was calculated by fitting the distribution of time differences between events, $\Delta t$, with an exponential function, 
\begin{equation}\label{eq:livetime_deltaT}
  N(\Delta t) = N_0 \exp(-\Delta t/\tau).
\end{equation}
This was done individually for each data taking run.
The selected runs have a total effective livetime of $T = \sum_{\text{runs } i}(N_i \cdot \tau_i) = (3274.0 \pm 1.9)\un{h}$, which corresponds to $89.4\un{\%}$ of the selected $153$ days period. 
The uncertainty on the livetime was included in the statistical error.

\section{Air shower reconstruction}\label{sec:reconstruction}
The energy of the primary particle cannot be measured directly, but has to be determined from the air shower parameters.
Properties of an air shower that are reconstructed by IceTop are the shower core position, its direction, and the shower size. 
The latter is a  measure of primary energy and is  defined as the signal $S_{\rm ref}$ measured at a certain distance $R_{\rm ref}$ from the shower axis. 
These properties are reconstructed by fitting the measured charges with a lateral distribution function and the signal times with a function describing the geometric shape of the shower front.

\subsection{The reference radius $R_\mathrm{ref}$}
The average logarithmic distance to the shower axis, $\langle\log R\rangle$, of signals participating in the fit for the given array configuration and energy range under investigation is about $125\un{m}$. 
While this number does depend on the primary energy and mass, it is limited by the relatively small size and the particular geometry of the 26-station array.
A constant $R_{\mathrm{ref}} = 125\un{m}$ was chosen in order to minimize the correlation between the parameters $S_\mathrm{ref}$ and $\beta$ in the fit. 
The shower size parameter is thus referred to as $S_{125}$.

\subsection{Time and charge distribution of air shower signals} \label{sub:expect_shower}
\paragraph{Lateral charge distribution}
IceTop tanks are not only sensitive to the number of charged particles, but also detect photons. 
Furthermore, the signal generated by a particle when it traverses the tank also depends on incident particle type, energy and direction.
Therefore, the charge expectation value in an IceTop tank at distance~$R$ from the shower axis was described by an empirical lateral distribution function found in Monte Carlo simulations~\cite{StefanThesis}:
\begin{equation}\label{eq:ldf}
  S(R) = S_{\mathrm{ref}} \cdot \left(\frac{R}{R_{\mathrm{ref}}}\right)^{-\beta -\kappa\, \log(R/R_{\mathrm{ref}})}.
\end{equation}
This is a second order polynomial in $\log R$ for the logarithm of the signal, $\log S(R)$:
\begin{equation}\label{eq:logldf}
  \log\, S(R) =  \log\, S_{\mathrm{ref}} - \beta\, \log\left(\frac{R}{R_{\mathrm{ref}}}\right) - \kappa\, \log^2\left(\frac{R}{R_{\mathrm{ref}}}\right).
\end{equation}
This function behaves unphysically at small distances to the shower axis~($R \lesssim 1\un{m}$). 
However, as described in the next subsection, all signals within~$11\un{m}$ of the core, are excluded from the fit. 
The free parameters of the function, in addition to the shower size,~$S_{\mathrm{ref}}$, are~$\beta$ and~$\kappa$, corresponding to the slope and curvature in the logarithmic representation at~$R=R_{\mathrm{ref}}$. 
The parameter~$\kappa$ is fixed at the average value of~$0.303$ found in simulation studies and it was verified that this constraint does not have a significant impact on the result.
Therefore, a fit of function~\eqref{eq:logldf} depends only on two explicit parameters~($S_{\mathrm{ref}}$,~$\beta$) and, since~$R$ depends on shower core position~($x_c,~y_c$) and direction~($\theta, \phi$), implicitly on four more.

\begin{figure*}
  \centering%
  \includegraphics[width=.5\textwidth]{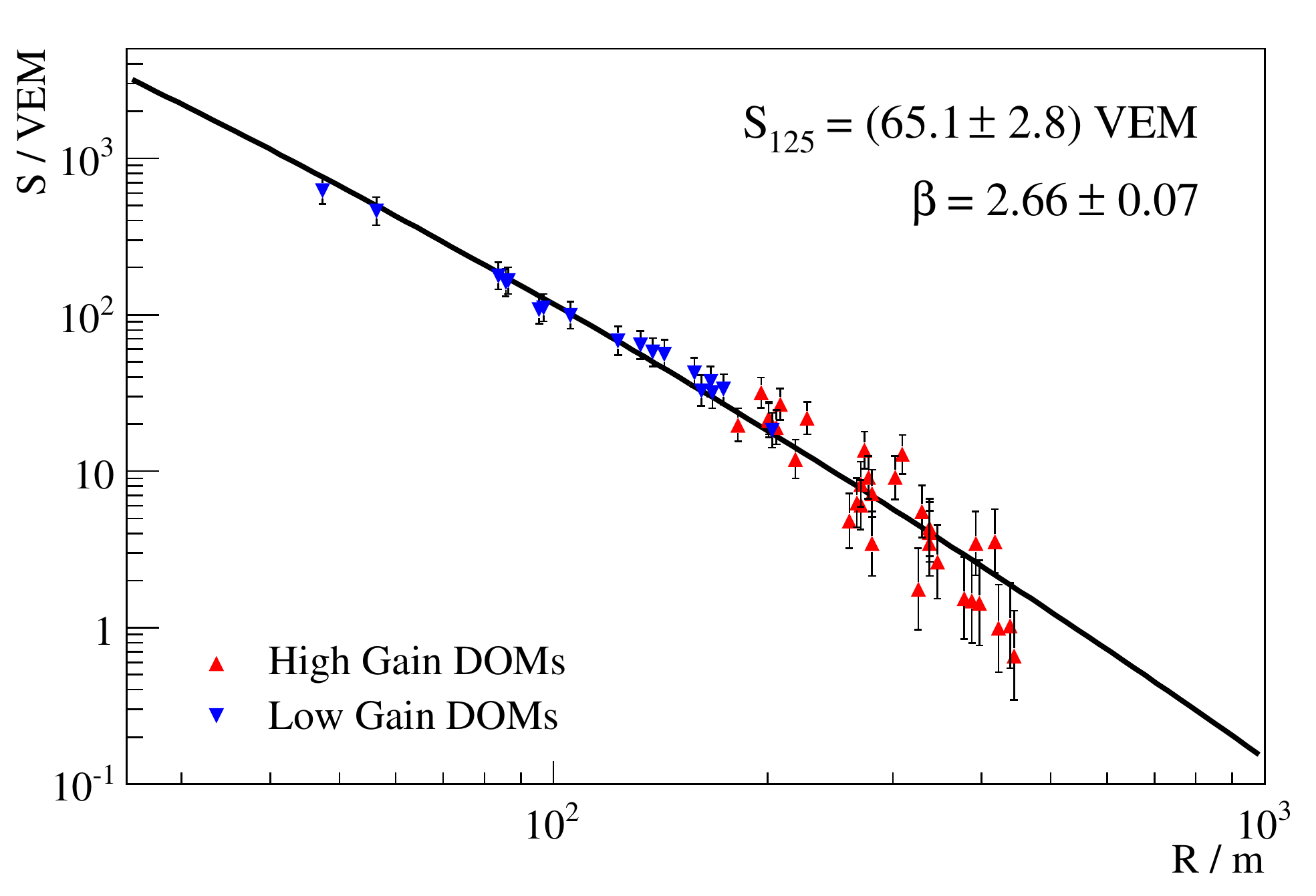}%
  \includegraphics[width=.5\textwidth]{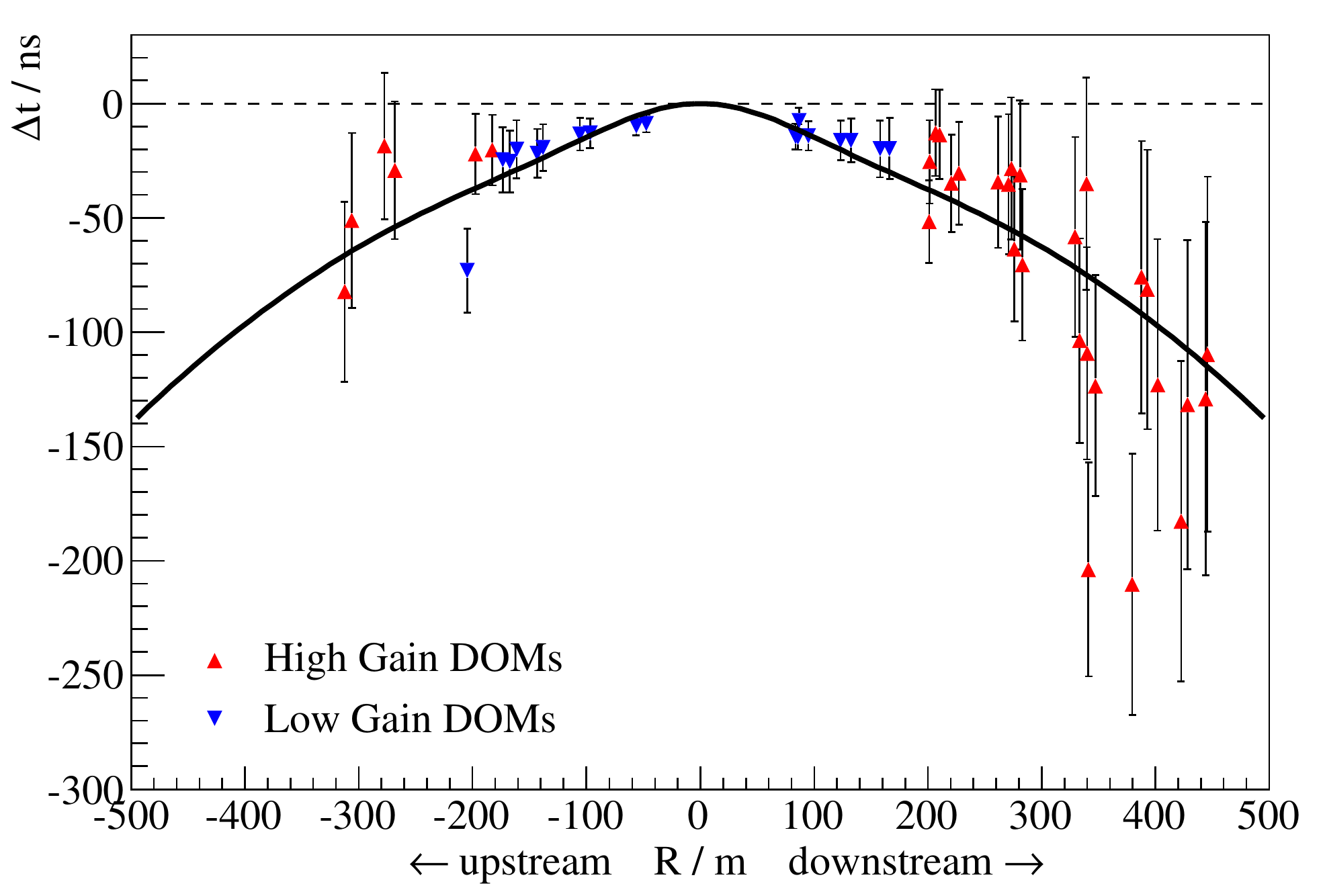}%
  \caption{Left: Example of an IceTop lateral fit. The shower triggered $25$ stations and the reconstructed shower size is $S_{125} = (65.1 \pm 2.8)\un{VEM}$. Right: Time residuals with respect to a plane perpendicular to the shower direction given by Eq.~\eqref{eq:curvature}. ``Upstream'' and ``downstream'' refer to tanks being hit before and after the shower core reaches the ground.}
  \label{fig:ldf}
\end{figure*}

In the following we will only refer to the reference radius of~$125\un{m}$ motivated in the previous subsection.
Figure~\ref{fig:ldf} shows an example of the lateral distribution function fit of a shower with $25$ triggered stations.

\paragraph{Time distribution}
The arrival times of the signals map out the shower front. 
The expected signal time of a tank at the position $\boldsymbol{x}$ was thus parametrized as
\begin{equation} \label{eq:time_expect}
  t(\boldsymbol{x}) = t_0 + \tfrac{1}{c}(\boldsymbol{x}_c - \boldsymbol{x}) \cdot \boldsymbol{n} + \Delta t(R).
\end{equation}
Here, $t_0$ is the time the shower core reaches the ground, $\boldsymbol{x}_c$ is the position of the shower core on the ground and $\boldsymbol{n}$ is the unit vector in the direction of movement of the shower. 
The ground was defined as the~\mbox{$\sqrt{S}$-weighted} average of participating tank altitudes, which varied by about~$3\un{m}$.
The term~$\Delta t(R)$ describes the shape of the shower front as a function of distance~$R$ to shower axis and is the time residual with respect to a plane perpendicular to the shower axis which contains~$\boldsymbol{x}_c$. 
Experimentally, the shower front can be described by the sum of a parabola and a Gaussian function, both symmetric around the shower axis:
\begin{equation}\label{eq:curvature}
  \Delta t(R) = a \, R^2 + b \left(\exp\left(-\frac{R^2}{2\sigma^2}\right) - 1\right),
\end{equation}
with the constants
 \[ a = 4.823 \, 10^{-4}\un{ns/m^2}, \qquad  b = -19.41\un{ns} , \qquad \sigma = 83.5\un{m}. \]

Function~\eqref{eq:time_expect} is fitted to the measured signal times with five free parameters: two for the core position, two for the shower direction and one for the reference time~$t_0$. 
Hence, the complete air shower reconstruction has the following parameters: position of the shower core~\mbox{$(x_c, y_c)$}, shower direction~$\theta$ and~$\phi$, shower size~$S_{125}$, slope parameter~$\beta$, and time at ground~$t_0$.

\subsection{Likelihood fit}\label{sub:likelihood_fit}

\paragraph{Likelihood function}
The functions~\eqref{eq:logldf}, \eqref{eq:time_expect} and \eqref{eq:curvature} describing the expectations for the charge and time of air shower signals were fitted to the measured data using the maximum likelihood method. 
In addition to terms for the signal charges and times, the likelihood function also takes into account stations that did not trigger so that the full likelihood function consisted of three factors. 
As usual we use the logarithm of the likelihood function:
\begin{equation}\label{eq:llh}
  \mathcal{L} = \mathcal{L}_q + \mathcal{L}_0 + \mathcal{L}_t.
\end{equation}

The first term,
\begin{equation}\label{eq:llh_q}
  \mathcal{L}_q = -\sum_i\frac{\bigl(\log S_i -
      \log S^{\rm fit}_i\bigr)^2}{2\,\sigma^2_q(S^{\rm fit}_i)} 
    - \sum_i\ln\bigl(\sigma_q(S^{\rm fit}_i)\bigr),
\end{equation}
describes the probability of measuring the charges $S_i$ if the fit expectation value at the position of the tank is $S^{\rm fit}_i$ as given by the lateral distribution function~\eqref{eq:ldf}. 
The sum runs over all tanks that have triggered. 
The signal fluctuations are described by a normal distribution of $\log S_i$ around $\log S^{\rm fit}_i$, with standard deviations $\sigma_q$ depending on the signal charge. 
The charge dependence of $\sigma_q$ has been determined experimentally from the local shower fluctuations between the two tanks of a station and are reasonably well reproduced by simulation~\cite{FabianThesis}. 
It can roughly be described by a linear improvement of~$\log(\sigma_q(\log S))$ until a saturation level is reached at~$S \approx 120\un{VEM}$.
The second sum in $\mathcal{L}_q$ accounts for the proper normalization of the signal likelihood and is required because the standard deviations $\sigma_q$ depend on the fitted signals.

The next term of the log-likelihood function~\eqref{eq:llh},
\begin{equation}\label{eq:llh_0}
  \mathcal{L}_0 = \sum_j\ln\Bigl(1-\bigl({P^{\rm hit}_j}\bigr)^2\Bigr),
\end{equation}
accounts for all stations $j$ that did not trigger. 
The probability that one tank in station $j$ delivers a signal at a given charge expectation value is
\begin{equation}
	\label{eq:p_hit}
	  P^\mathrm{hit}_j = \frac{1}{\sqrt{2\pi}\sigma_q(S^\mathrm{fit}_j)} \cdot
	  \int\limits_{\log S^\mathrm{thr}_j}^{\infty}
	    \exp\left(-\frac{\bigl(\log S_j-\log S^\mathrm{fit}_j \bigr)^2}{2\sigma^2_q(S^\mathrm{fit}_j)}\right)\dd \log S_j.
\end{equation}
The lower integration limit is defined through the charge threshold of $S^\mathrm{thr}_j = 0.3\un{VEM}$ for the tank signal, as determined by the retriggering procedure described in Section~\ref{sub:retriggering}.
The charge expectation value, $S^\mathrm{fit}_j$, was evaluated for the center of a line joining the centres of the two tanks.
Since the two tanks of one station are operated in coincidence, there are no single untriggered tanks. 
Equation~\eqref{eq:llh_0} is an approximation because it assumes that $P_j^\mathrm{hit}$ in the two tanks is independent.
Of course, there is a natural correlation in the signal expectation values of two nearby tanks because they have a similar value of the lateral distribution function. 
However, the fluctuations about this expectation value are assumed to be uncorrelated. 

The third term of function~\eqref{eq:llh}, $\mathcal{L}_t$, describes the probability for the measured set of signal times,
\begin{equation}
	\label{eq:llh_time}
	  \mathcal{L}_t= -\sum_i \frac{(t_i - t_i^{\rm fit})^2}{2\,\sigma_t^2(R_i)} - \sum_i\ln(\sigma_t(R_i)/\mathrm{ns}),
\end{equation}
where the index $i$ runs over all tanks, $t_i$ is the measured signal time of tank $i$ and $t_i^{\rm fit}=t(\boldsymbol{x}_{i})$ is the fitted expectation value according to function~\eqref{eq:time_expect}. 
The arrival time fluctuations $\sigma_t(R_i)$ depend on the distance $R_i$ of tank $i$ to the shower axis, and are the RMS of the arrival time distribution found in experimental data~\cite{FabianThesis}.

\paragraph{Fit procedure}
The likelihood fit was seeded with first-guess calculations for the core and the direction of the shower. 
As a first estimate of the core position the centre-of-gravity of tank positions $\boldsymbol{x}_i$ weighted with the square root of the charges was calculated:
\begin{equation}\label{eq:cog}
  \boldsymbol{x}_{\rm COG} = \frac{\sum_{i}\sqrt{S_i}\,\boldsymbol{x}_i}{\sum_{i}\sqrt{S_i}}.
\end{equation}
The square root of $S$ used as a weight was chosen based on a study of the achievable fit accuracy.
The starting values for shower direction and arrival time were obtained by fitting a plane to the signal times. 

The likelihood minimisation is then done in several iterations to improve the stability of the fit. 
At first the shower direction is fixed and only the lateral fit of the charges is iterated with the free parameters $S_{125}$, $\beta$, and core position. 
After each iteration, those tanks that are closer than $11\un{m}$ to the shower axis are removed from the fit. 
Iteration is stopped when no more tanks are removed from the fit.
The reason for this step was that very large signals tended to unnaturally attract shower cores, which had a negative effect on the shower core resolution in the vicinity of stations.
Additionally, this mitigated the effect of saturated pulses.
Then, a final iteration is done in which description of the shower curvature is included and the shower direction is varied.

\section{Simulation of air showers and the IceTop detector} \label{sec:simulation}
The relation between the measured signals and the energy of the primary particle, as well as detection efficiency and energy resolution were obtained from CORSIKA~\cite{CORSIKA} air shower simulations and simulations of the IceTop detector. 

\subsection{Air shower simulation}\label{sub:corsika}
We simulate the development of air showers in the atmosphere using the simulation code CORSIKA~\cite{CORSIKA}.
Inside CORSIKA, the hadronic component of the air showers was simulated using the models SIBYLL2.1~\cite{SIBYLL1,SIBYLL2} and FLUKA~2008.3~\cite{Fluka1,Fluka2} for the high and low energy interactions, respectively. 
The electromagnetic component was simulated using the EGS4 code~\cite{EGS4} and no `thinning' (reduction of the number of traced particles) was applied. 
To study systematic effects of the hadronic interaction model, small samples of showers were simulated using the QGSJET-II~\cite{QGSJET1,QGSJET2} and EPOS~1.99~\cite{Epos} high energy interaction model. 
Two different parameterizations of the South Pole atmosphere from two days in 1997 based on the MSIS-90-E model~\cite{MSIS} were used: July 1st and October 1st (CORSIKA atmospheres 12 and 13). 
The July atmosphere has a total overburden of $692.9\un{g/cm^2}$, while the October atmosphere has an overburden of $704.4\un{g/cm^2}$.
The July atmosphere was used in the data analysis, because its total overburden is close to the average measured overburden of $695.5\un{g/cm^2}$ and its profile corresponds to that of a South Pole winter atmosphere.
The October atmosphere model was used to study systematic uncertainties due to the atmospheric profile used in the simulation.

\subsection{Detector simulation}\label{sub:detsim}
The output of the CORSIKA program, i.\,e.~the shower particle types, positions and momenta at the observation level of $2835\un{m}$, were injected into the IceTop detector simulation. 
The simulation determines the amount of light produced by the shower particles in the tanks followed by the simulation of the PMT, the DOM electronics and the trigger chain.

The Cherenkov emission inside the tanks is simulated using Geant4~\cite{Geant4_1,Geant4_2}. 
All structures of the tank, the surrounding snow, including individual snow heights on top of each tank, as well as the air above the snow are modeled realistically~\cite{ThomasThesis}. 
The snow heights used in the simulation corresponded to those measured in January 2007 (see Fig.~\ref{fig:snowheights}). 
In order to save computing time, Cherenkov photons are not tracked; only the number of photons emitted in the wavelength interval $300\un{nm}$ to $650\un{nm}$ is recorded. 
Using Geant4 simulations, that include Cherenkov photon tracking until photons reach the PMT, it was shown that the number of detected photons scales linearly with the number of emitted photons, independent of incident particle type and energy. 
The propagation of Cherenkov photons is modeled by distributing the arrival times according to an exponential distribution, which is tuned such that simulated waveform decay times match those observed in experimental data ($26.5\un{ns}$ for zirconium lined tanks and $42.0\un{ns}$ for tanks with Tyvek bag).

The number of photoelectrons corresponding to $1\un{VEM}$ was taken from the VEM calibration of the real tanks and used as an input for the simulation.
The simulated tanks were then calibrated by generating muon spectra as in experimental data using air shower simulations with primary energies between $3\un{GeV}$ and $30\un{TeV}$ and zenith angles up to $65\degrees$.
Thus, the ratio between the number of emitted Cherenkov photons and observed photoelectrons was determined by the VEM calibration of simulated tanks.

In the next step the generated photoelectrons are injected into a detailed simulation of the PMT followed by the analog and digital electronics of the DOM. 
To simulate the photomultipliers, Gaussian single photoelectron waveforms with a random charge according to the average single photoelectron spectrum are superimposed~\cite{PMTPaper}.
Afterwards, a saturation function is applied to the resulting waveforms.
In the DOM simulation, the pulse shaping due to the analog front end electronics is applied to the output of the PMT simulation. 
This includes the individual shaping of the signal paths to the ATWD and the discriminators, as well as the simulation of the droop effect induced by the toroid that couples the high voltage circuits of the PMT to the readout electronics. 
Then, the discriminators are simulated and the local coincidence conditions are evaluated. 
Finally, the waveform digitization and the array trigger are simulated.

Simulated data are of the same format as the experimental data and were reconstructed in the same way, as described in the previous section.

\subsection{Simulation datasets}\label{sub:simulation_datasets}
In this analysis we describe the cosmic ray composition just with the two extreme elements hydrogen and iron. 
The justification comes from the fact that the final result is not sensitive to details of the composition but only to the mean logarithmic mass.

In total $2 \cdot 10^5$ showers of proton and iron primaries in the energy range between $100\un{TeV}$ and $100\un{PeV}$ were generated in~30~logarithmic energy bins according to an $E^{-1}$ spectrum.
For the analysis, the events are reweighted to an $E^{-3}$ flux, which is closer to the results of previous experiments and thus reduces systematic biases (see also Section~\ref{sub:systematics-response}).
In addition to pure proton and iron simulations we also combined the datasets using a parametrization of Glasstetter's two-component model~\cite{glasstetter}.
We transformed the proton flux to the form
\begin{equation}\label{eq:knee}
  \frac{\dd I}{\dd\ln E} = I_0 
    \left(\frac{E}{1\un{PeV}}\right)^{\gamma_1+1} 
    \left(1 + \left(\frac{E}{E_\mathrm{knee}}\right)^{\eps}
      \right)^{(\gamma_2-\gamma_1)/\eps},
\end{equation}
as suggested in~\cite{TerAntonyanFlux}, with $I_0 = 3.89 \cdot 10^{-6}\un{m^{-2}s^{-1}sr^{-1}}$, $\gamma_1 = -2.67$, $\gamma_2 = -3.39$, $E_\mathrm{knee} = 4.1\un{PeV}$, and $\eps = 2.1$. The iron flux was used as specified in Ref.~\cite{glasstetter}:
\begin{equation}
  \frac{\dd I}{\dd\ln E} = 1.95 \cdot 10^{-6}\un{m^{-2}s^{-1}sr^{-1}} \cdot \left(\frac{E}{1\un{PeV}}\right)^{-1.69}.
\end{equation}
The total flux was then normalized to the same $E^{-3}$ spectrum as in case of the single component Monte Carlo.

Since shower generation is CPU intensive the same showers were sampled several times inside a circle with a radius of $1200\un{m}$ around the center of the 26 station IceTop array. 
The number of samples was chosen for different energy bins such that every shower would remain on average only once in the final sample after applying the cuts described in the next section. 
This ensures a good balance between an effective use of the generated showers and the artificial fluctuations introduced by oversampling.

\begin{figure}
  \centering
  \includegraphics[width=.5\columnwidth]{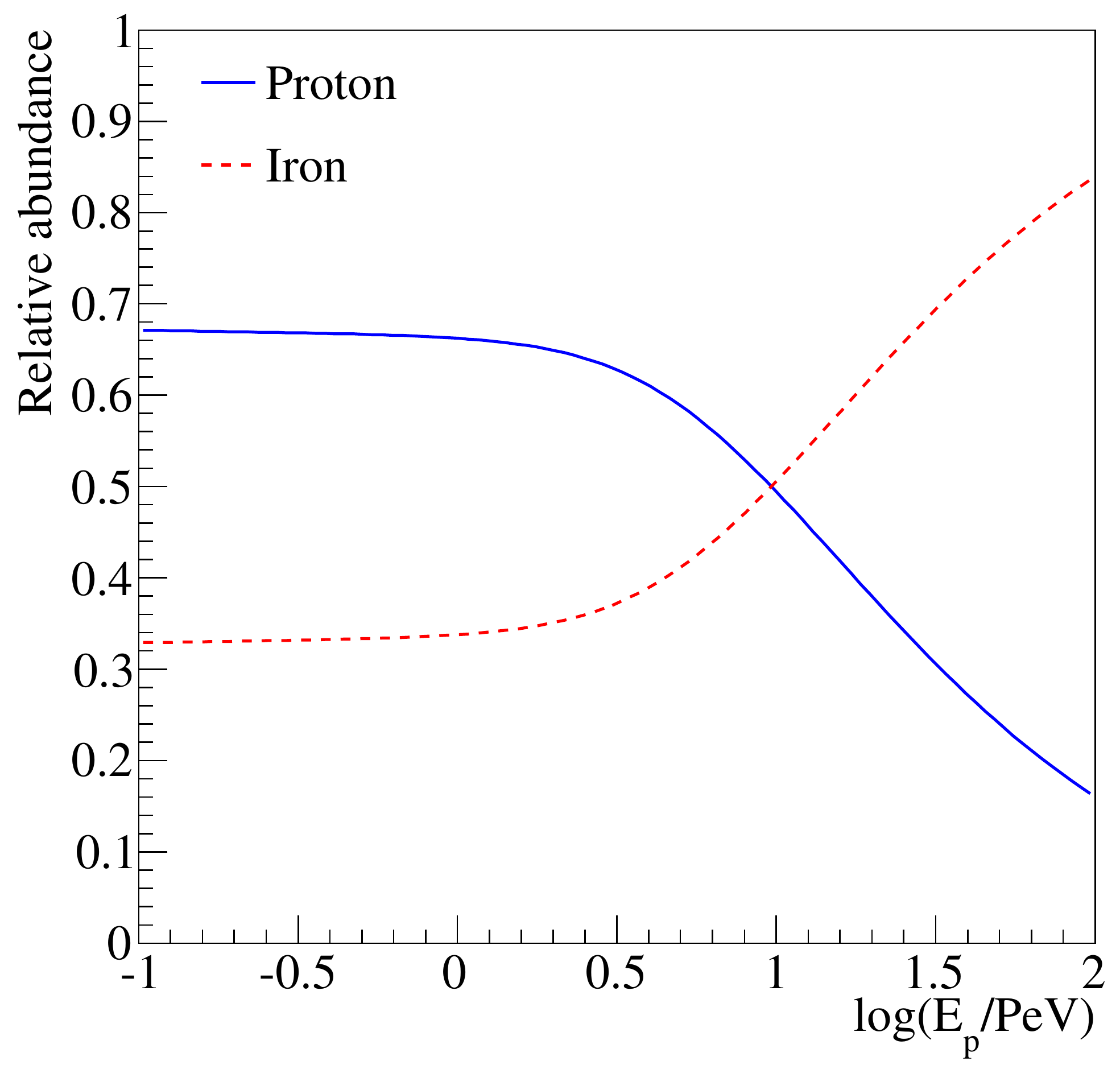}
  \caption{Relative abundance of proton and iron in our parametrization of Glasstetter's two-component model as a function of primary energy. Above about $10\un{PeV}$ the spectrum is dominated by iron.}
  \label{fig:twocomponent}
\end{figure}

\section{Event selection and reconstruction performance} \label{sub:final_selection}
\paragraph{Quality cuts}
Based on the reconstruction results the following quality criteria were required for each event entering the final event sample, for both simulated and experimental data: 
\begin{itemize}
 \item Containment cut: The reconstructed core and the first-guess core position had to be at least $50\un{m}$ inside the boundary of the array. The array boundary is defined by the polygon with vertices at the centers of stations at the periphery of the array and edges connecting these stations. This cut defines a fiducial area of $A_{\rm cut} = 0.122\un{km^2}$. Furthermore, it was required that the station containing the largest signal is not on the border of the array.
 \item Only events with zenith angles~$\theta < 46\degrees$ were considered.
 \item The reconstruction uncertainty on the core position had to fulfill $\sigma_\mathrm{core} = \sqrt{\sigma_x^2 + \sigma_y^2} < 20\un{m}$.
 \item The slope parameter $\beta$ had to be in the range $2.0 \leq \beta < 4.5$ because most events with $\beta$ values outside this range were badly reconstructed and because $\beta$ was limited in the fit. The removed events had predominantly low primary energies, $E_0 \lesssim 1\un{PeV}$.
\end{itemize}

In the experimental dataset $3\,096\,334$ events passed the quality cuts.
Passing rates for the individual cuts are shown in Table~\ref{tab:cutflow} for events with $S_{125} > 1\un{VEM}$. 
Differences between data and Monte Carlo are discussed later in Section~\ref{sub:systematics-cuts}.

\begin{table*}
  \centering
  \caption{Passing rates of the quality cuts described in the text for events with $S_{125} > 1\un{VEM}$. Statistical errors on experimental data are negligible.}
  \vspace*{1ex}
  \begin{tabular}{lrrrr}
    \toprule
    \multicolumn{1}{c}{\multirow{2}{*}{\textbf{Cut}}}
      & \multicolumn{2}{c}{\textbf{Experimental data}} & \multicolumn{2}{c}{\textbf{Monte Carlo}} \\
                                       & Passing rate & Cumulative & Passing rate           & Cumulative           \\
    \midrule
    $N_\mathrm{station} > 5$ and $S_{125} > 1\un{VEM}$
                                       & $100\%$      &            & $100\%$                &                      \\
    Largest signal contained           & $42.5\%$     & $42.5\%$   & $(39.4 \pm 0.5)\%$     & $(39.4 \pm 0.5)\%$   \\
    First guess core contained         & $95.8\%$     & $40.7\%$   & $(95.4 \pm 0.4)\%$     & $(37.6 \pm 0.5)\%$   \\
    Core contained                     & $78.9\%$     & $32.1\%$   & $(81.1 \pm 0.5)\%$     & $(30.5 \pm 0.6)\%$   \\
    Zenith $\theta < 46\degrees$       & $96.3\%$     & $30.9\%$   & $(96.4 \pm 0.5)\%$     & $(29.4 \pm 0.6)\%$   \\
    $\sigma_\mathrm{core} < 20\un{m}$  & $99.7\%$     & $30.8\%$   & $100\%$                & $(29.4 \pm 0.6)\%$   \\
    $2.0 \leq \beta < 4.5$             & $98.1\%$     & $30.2\%$   & $(99.7 \pm 0.1)\%$     & $(29.3 \pm 0.6)\%$   \\
    \bottomrule
  \end{tabular}
  \label{tab:cutflow}
\end{table*}

\begin{figure*}
  \subfigcapskip-2ex%
  \hfill%
  \subfigure[Core resolution]{%
    \includegraphics[width=0.5\textwidth]{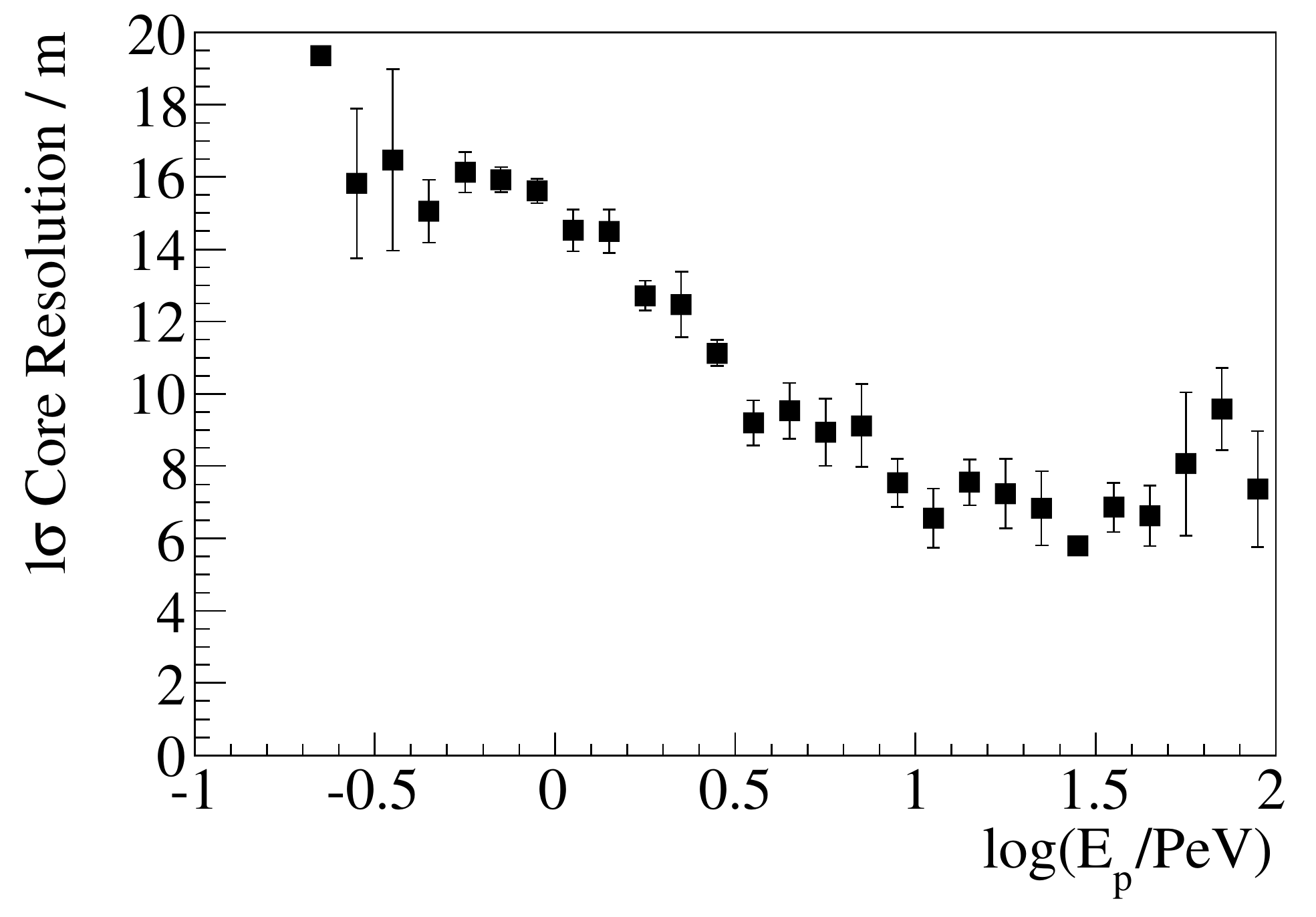}%
    \label{fig:core_resolution}}%
  \hfill%
  \subfigure[Angular resolution]{%
    \includegraphics[width=0.5\textwidth]{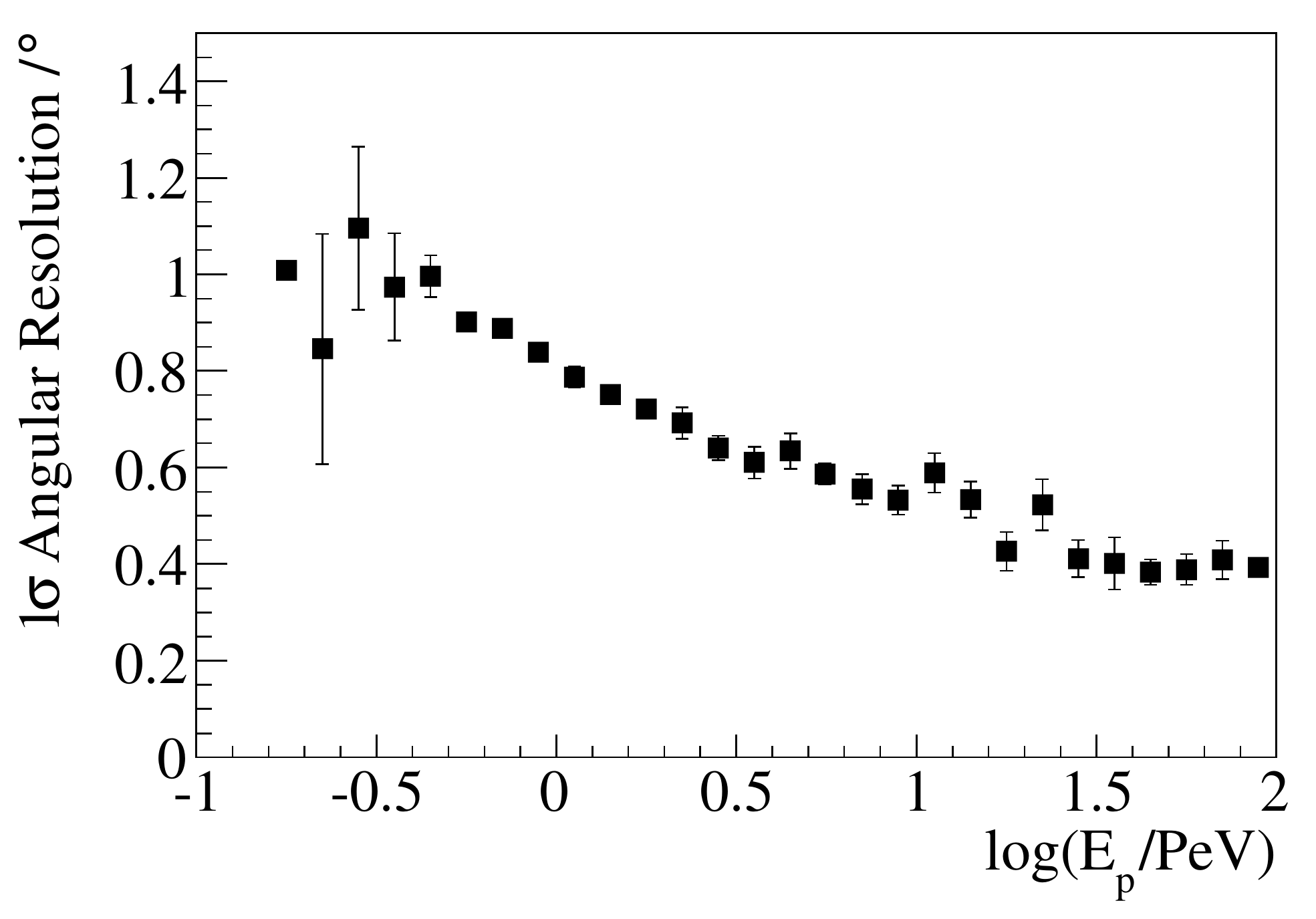}%
    \label{fig:angular_resolution}}%
  \hfill%
  \caption{Core position and angular resolution for showers with $\theta \leq 30\degrees$, obtained from the two-component Monte Carlo. At high energies, the distance between true and reconstructed core position of 68\% of showers is $7\un{m}$ or less. The angle between true and reconstructed directions of 68\% of showers at $1\un{PeV}$ is smaller than $0.8\degrees$ and this value decreases to $0.4\degrees$ at $100\un{PeV}$.}
  \label{fig:core_and_angle_resolution}
\end{figure*}

\paragraph{Reconstruction performance}
Core position and angular resolution, shown in Fig.~\ref{fig:core_and_angle_resolution}, are key criteria for the performance of air shower reconstruction.
The~$1\sigma$ core resolution is defined as the~$68\%$ quantile of the cumulative distribution of the distances between true and reconstructed shower cores; correspondingly the angular resolution is defined as the angle between true and reconstructed shower direction.
The numbers shown are for showers with zenith angle $\theta \leq 30\degrees$, obtained from the two-component Monte Carlo after applying the quality cuts listed in the previous paragraph. 
At the highest energies, a core resolution of $7\un{m}$ and an angular resolution of $0.4\degrees$ were achieved.
In the most inclined zenith angle range considere in this analysis, $40\degrees \leq \theta < 46\degrees$, a core resolution of $10\un{m}$ and an angular resolution of $0.5\degrees$ was achieved.

\section{Determination of energy spectra}\label{sec:energy_reco}

\begin{figure*}
  \centering%
  \includegraphics[width=.5\textwidth]{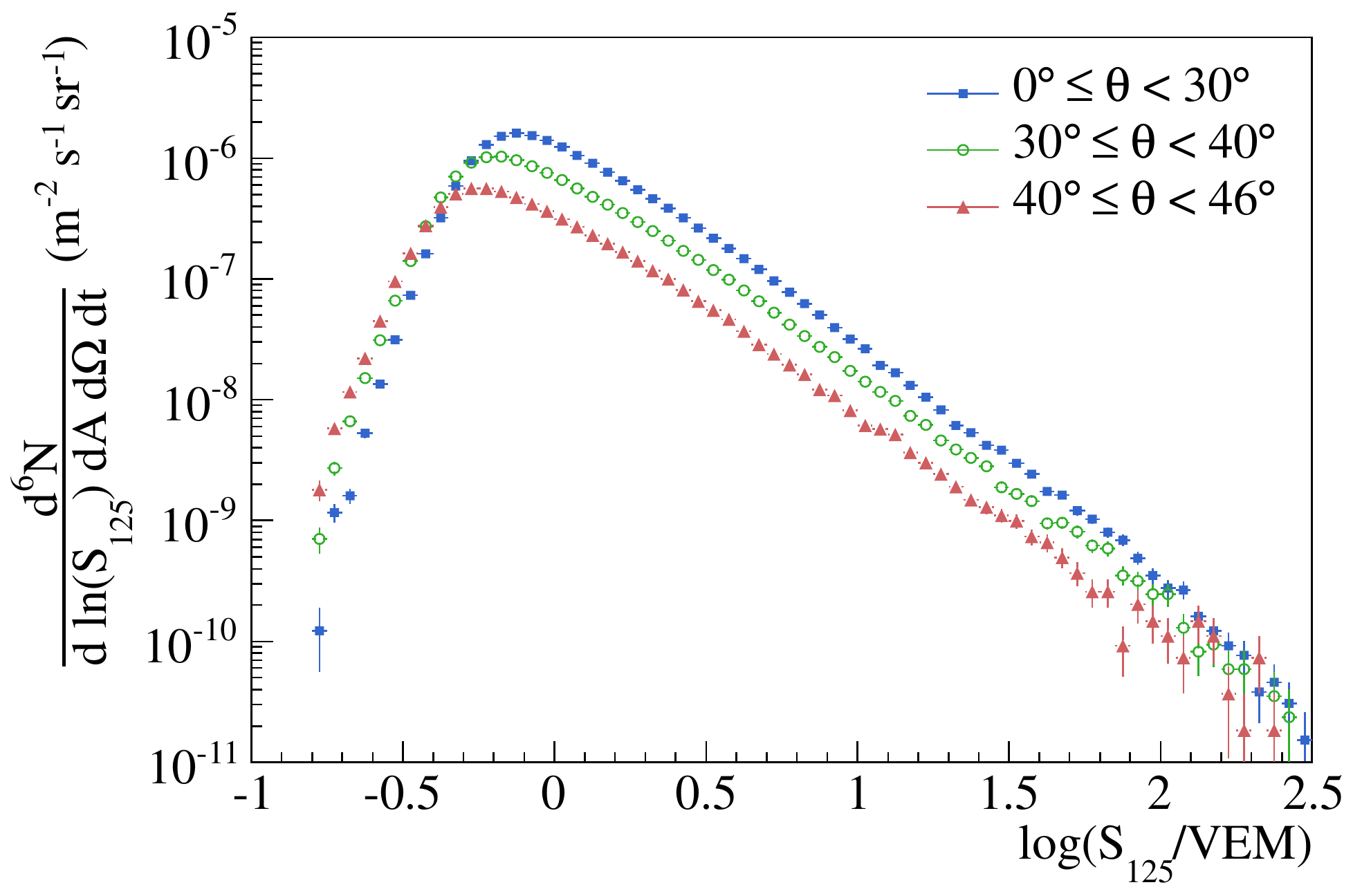}%
  \includegraphics[width=.5\textwidth]{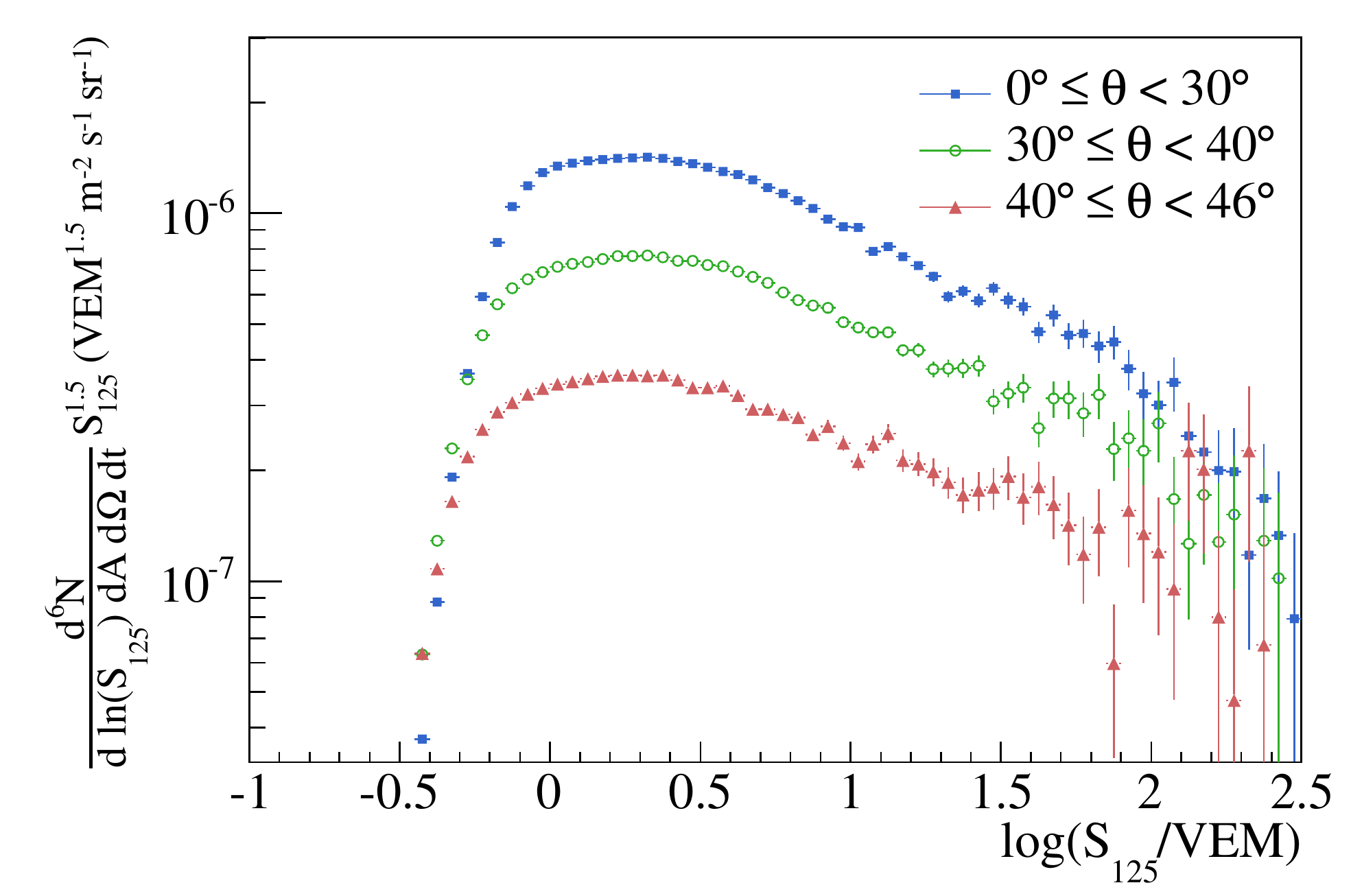}%
  \caption{Left: Reconstructed shower size spectra in three zenith angle bins. The energy spectrum was derived from these spectra using an unfolding method as described in Section~\ref{sec:energy_reco}. On the right, the same spectra are shown, weighted with $S_{125}^{1.5}$. In this representation it is clearly visible that the spectra are not pure power laws, but there is a clear structure above $\log S_{125} \approx 1.4$.}
  \label{fig:shower_size_spectra}
\end{figure*}

Using the reconstruction methods and quality cuts described in Sections~\ref{sec:reconstruction} and~\ref{sub:final_selection}, the shower size spectra shown in Fig.~\ref{fig:shower_size_spectra} were obtained.
In this analysis, the data were split into three zenith angle ranges roughly equidistant in $\sec\theta$, defined as:
\begin{equation}\label{eq:zenith_ranges}
  \Omega_1 = \left[0\degrees , 30\degrees\right],\qquad
  \Omega_2 = \left[30\degrees , 40\degrees\right],\qquad
  \Omega_3 = \left[40\degrees , 46\degrees\right] .
\end{equation}
A steepening of the spectral slope is visible at $\log(S_{125}/\mathrm{VEM}) = 0.5$ and a possible flattening at about $\log(S_{125}/\mathrm{VEM}) = 1.4$.
To determine the energy spectrum from measured data, these $S_{125}$ spectra were unfolded. 
Unfolding was performed for each zenith angle range independently.

\subsection{General method}
For the unfolding procedure the response of the detector to a primary particle of mass~$M$, energy~$E_p$, zenith angle~$\theta$, azimuth~$\phi$, and core position~\mbox{$(x_c, y_c)$} has to be determined from simulation. 
In this analysis we consider only an unfolding of energies.
Within each zenith angle range, we average over the dependencies on zenith, azimuth, and core position.
The response of the detector is the probability of measuring a shower size $S_{125}$ given a primary energy $E_p$ and mass $M$ in a certain zenith range $\Omega_k$.

In a discrete formulation we define a response matrix $\mathbf{R}$ which relates the bin contents $N^s_i\ (i=1,\ldots,m)$ of a measured $S_{125}$ spectrum with the bin contents $N^e_j\ (j=1,\ldots,n)$ of a primary energy spectrum for a fixed zenith range $\Omega_k$:
\begin{equation}\label{eq:folding}
	N^s_i = R_{ij}^{(k)} \, N^e_j.
\end{equation}
The response matrix elements $R_{ij}^{(k)}$ are defined as acceptance integrals
\begin{equation}\label{eq:Rijk}
  R_{ij}^{(k)} = \frac{
    \sum\limits_M 
     \int\limits_{\Delta E_p^j} \!\!\!\dd E_p 
    \, \int \dd\Omega
    \, \int\! \dd A_\bot
      \ \Phi_M(E_p) \ p_M^{(k)}(S_{125}^i \,|\, E_p)
  }{
    \rule{0pt}{9pt}\sum\limits_M 
    \int\limits_{\Delta E_p^j} \!\!\! \dd E_p 
    \int\limits_{\Omega_k} \!\dd\Omega
     \int\limits_{A_\mathrm{cut}} \!\!\!\dd A_\bot
      \ \Phi_M(E_p)
  }.
\end{equation}
The model flux $\Phi_M(E_p)$ of nuclei with mass $M$ weighted by their acceptance function
\begin{equation}
  p_M^{(k)}(S_{125}^i \,|\, E_p) = p(S_{125}, \Omega_k \,|\, E_p, x_c, y_c, \theta, \phi; M) 
\end{equation}
is integrated over primary energy bin~$E_p^j$, the angles~$\theta$ and~$\phi$, and area $A_\bot$ projected on a plane perpendicular to the particle direction.
It is summed over all mass components~$M$ that contribute to the assumed composition model. 
$R_{ij}^{(k)}$ is normalized to the flux integrated over bin~$j$ in~$E_p$, solid angle~$\Omega_k$, and fiducial area~$A_\mathrm{cut}$.
The function~$p_M$ is the probability of an event with mass~$M$ and kinematical variables~$(E_p, x_c, y_c, \theta, \phi)$ to be reconstructed with shower size~$S_{125}^i$ in bin~$i$ and zenith angle~$\theta$ in the range~$\Omega_k$, and to pass all cuts listed in Section~\ref{sub:final_selection}. 
Thus,~$R_{ij}^{(k)}$ for a given primary energy bin~$j$, is the ratio between number of events measured in~$S_{125}$ bin~$i$ and zenith bin~$k$, that pass all cuts, and the true number of events in that energy bin~$j$ and zenith bin~$k$ inside the fiducial area.
Since the~$E_p$~bins of~$R_{ij}^{(k)}$ are independent, the total flux model only affects weighting of events within one bin, but not neighboring bins.
The flux normalization in~$R_{ik}^{(k)}$ cancels out, and the dependence on the spectral index of the flux model is small (see also Section~\ref{sub:systematics-response}).
The integrals in Eq.~\eqref{eq:Rijk} were determined numerically using the Monte Carlo method. 

With the normalisation to the full flux integral the response matrix has the following normalisation properties (we drop the superscript $k$ for zenith range):
\begin{equation}\label{eq:R_normalisation}
  \sum_i R_{ij} = \varepsilon_j, \qquad \sum_j R_{ij} = 1.
\end{equation}
That means, for a given energy bin~$j$ the sum of the probabilities to be detected in any signal bin is the efficiency~$\eps_j$; for a given $S_{125}$ bin $i$ the probability to belong to any energy $E_p$ is unity.
The efficiency depends on the energies, the core position $(x_c, y_c)$ and the angles. 

To obtain the primary energy spectrum from the measured signals the matrix equation~\eqref{eq:folding} has to be inverted: 
\begin{equation}\label{eq:unfolding}
  N^e_j = \left( R^{-1}\right)_{ji} \, N^s_i.
\end{equation}
For this unfolding procedure we use an iterative algorithm, which properly accounts for the statistical fluctuations as will be described Section~\ref{sec:unfolding}.

\subsection{Evaluation of response matrices, efficiencies and resolutions} \label{sec:response_matrix}

\begin{figure*}
  \centering%
  \includegraphics[width=.5\textwidth]{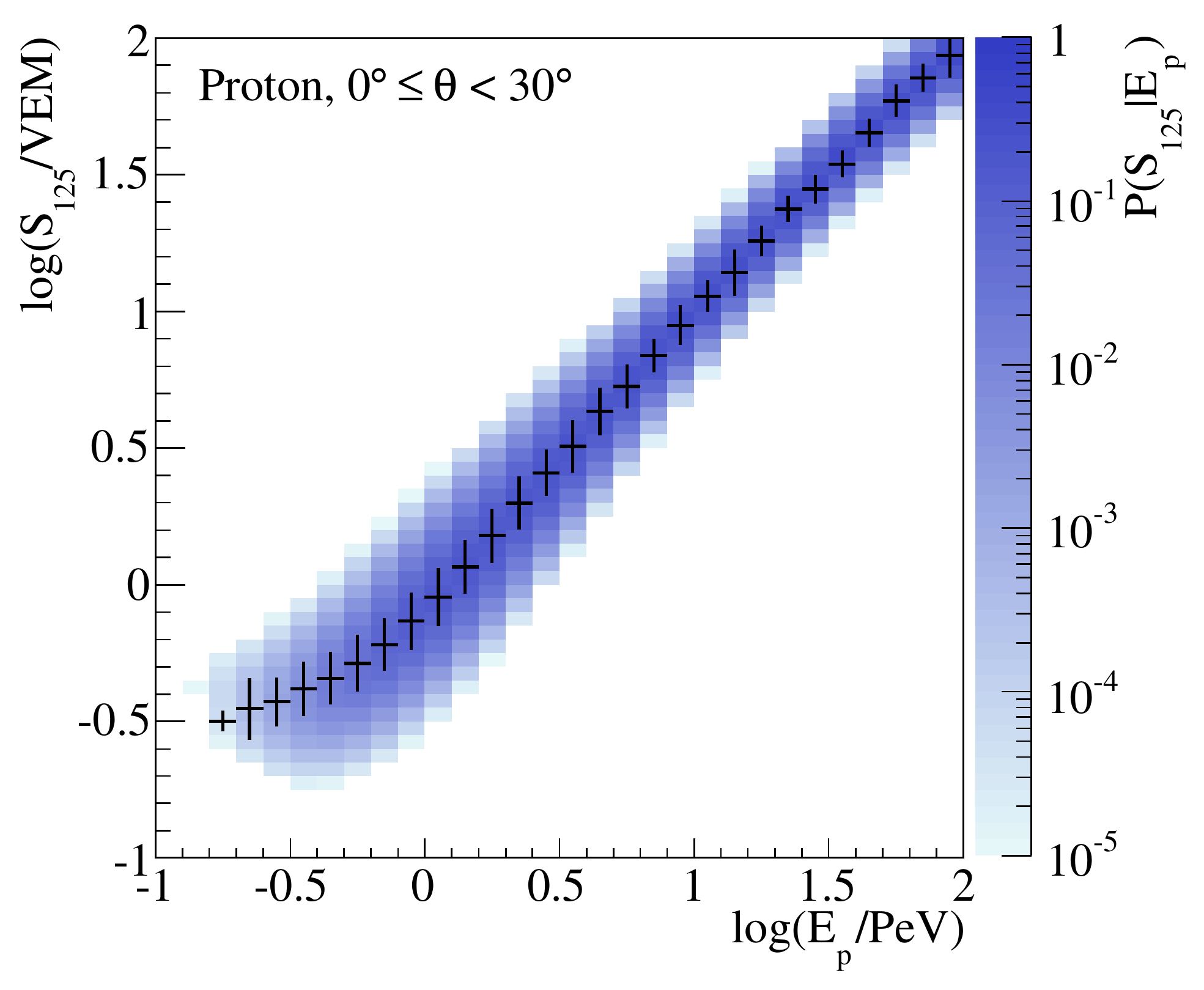}%
  \includegraphics[width=.5\textwidth]{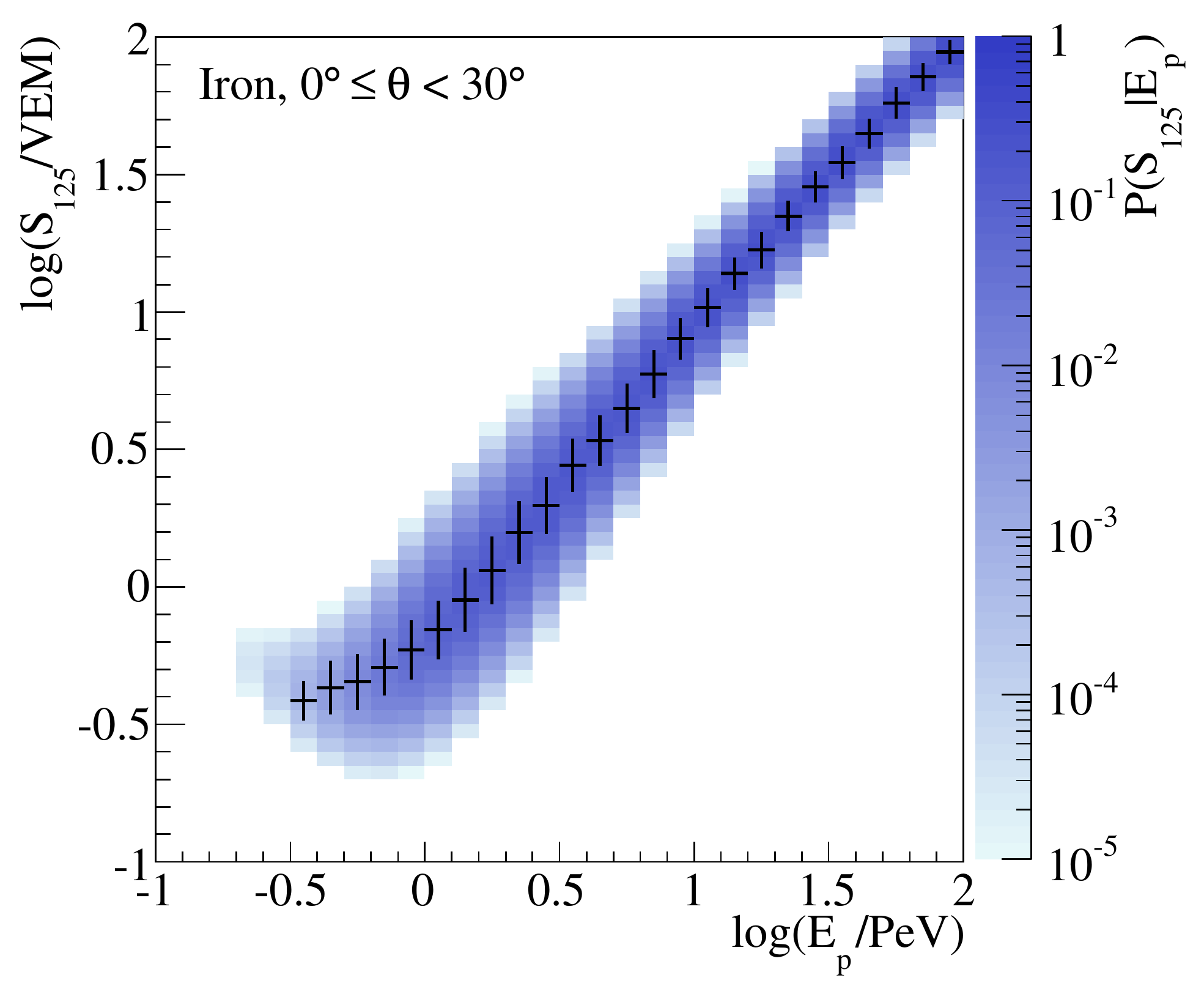}%
  \caption{Response matrix: shower size $S_{125}$ distribution as a function of primary energy for simulated proton (left) and iron (right) showers with zenith angles up to $30\degrees$. The crosses give the mean value and spread (RMS) of the distribution in each energy bin.}
  \label{fig:s125_vs_energy_proton}
\end{figure*}

\begin{figure}
  \centering
  \subfigcapskip-2ex
  \subfigure[]{
    \includegraphics[width=.6\columnwidth]{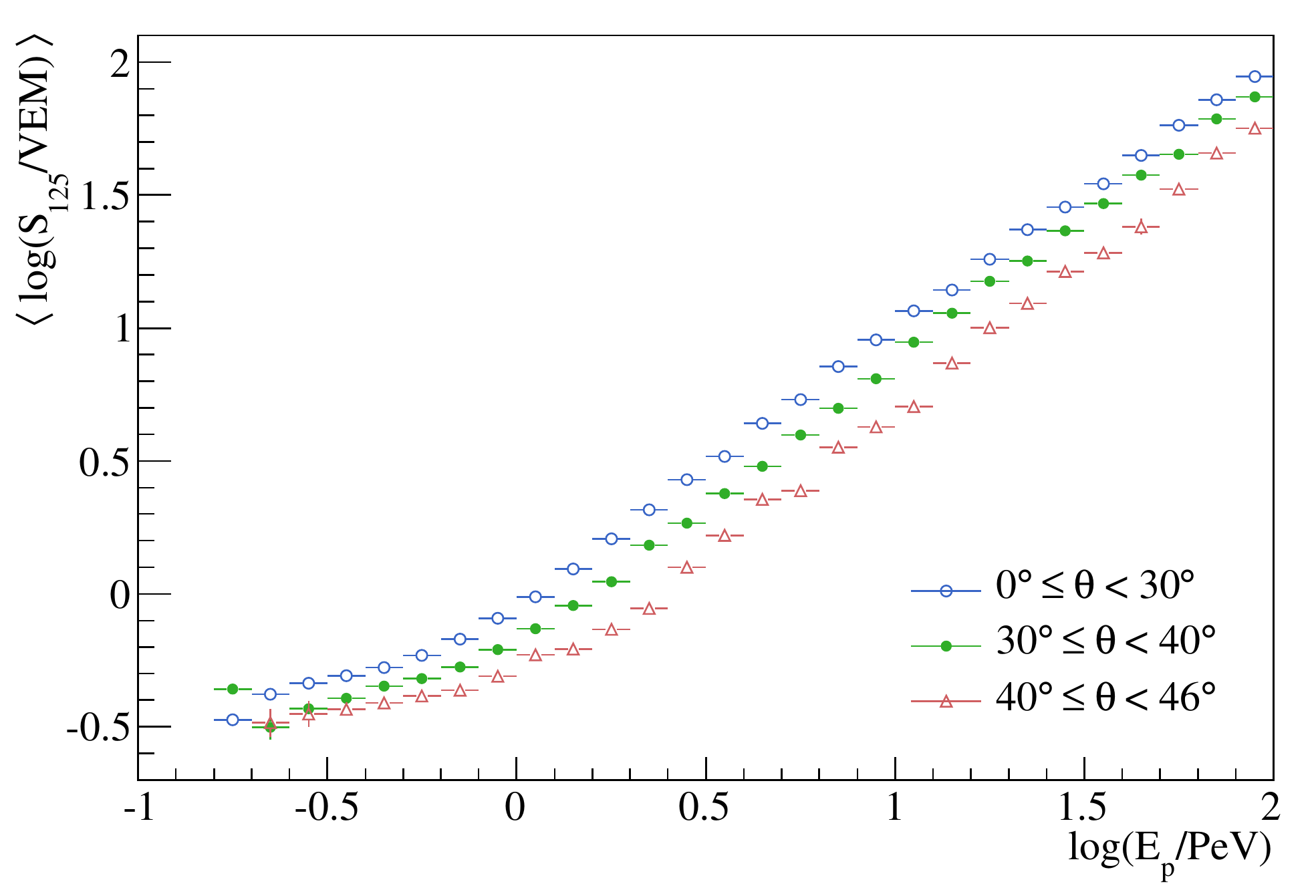}
    \label{fig:s125_vs_energy_proton_profiles}}
  \\
  \subfigure[]{
    \includegraphics[width=.6\columnwidth]{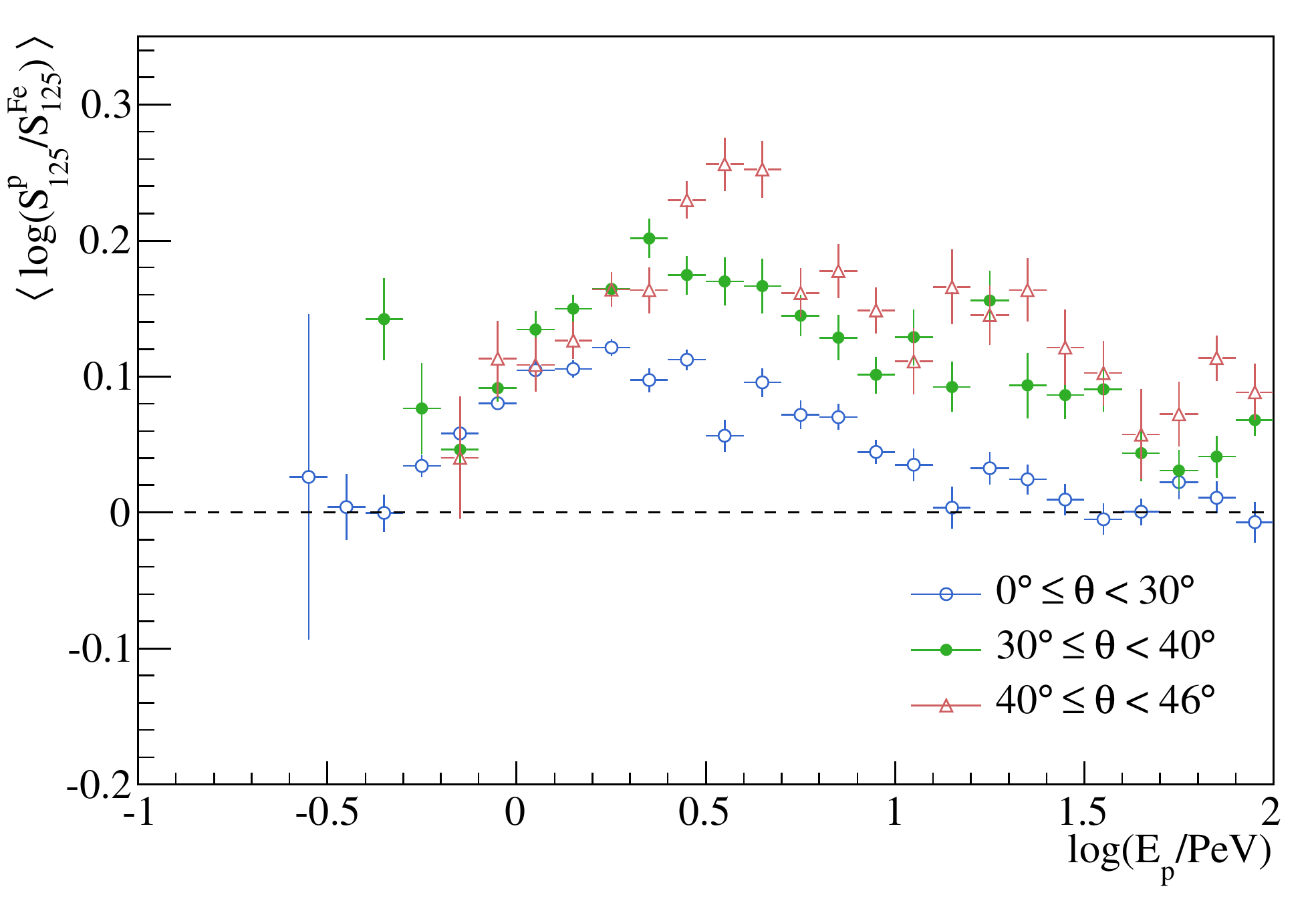}
    \label{fig:s125_vs_energy_p-fe}}
  \caption{\subref{fig:s125_vs_energy_proton_profiles} Mean shower size as a function of energy for proton showers of various zenith angles. 
  \subref{fig:s125_vs_energy_p-fe} Shower size ratio between proton and iron showers. The ratio increases for larger zenith angles because the attenuation of iron showers is stronger than for proton showers.}
\end{figure}

\begin{figure}
  \centering
  \includegraphics[width=.6\columnwidth]{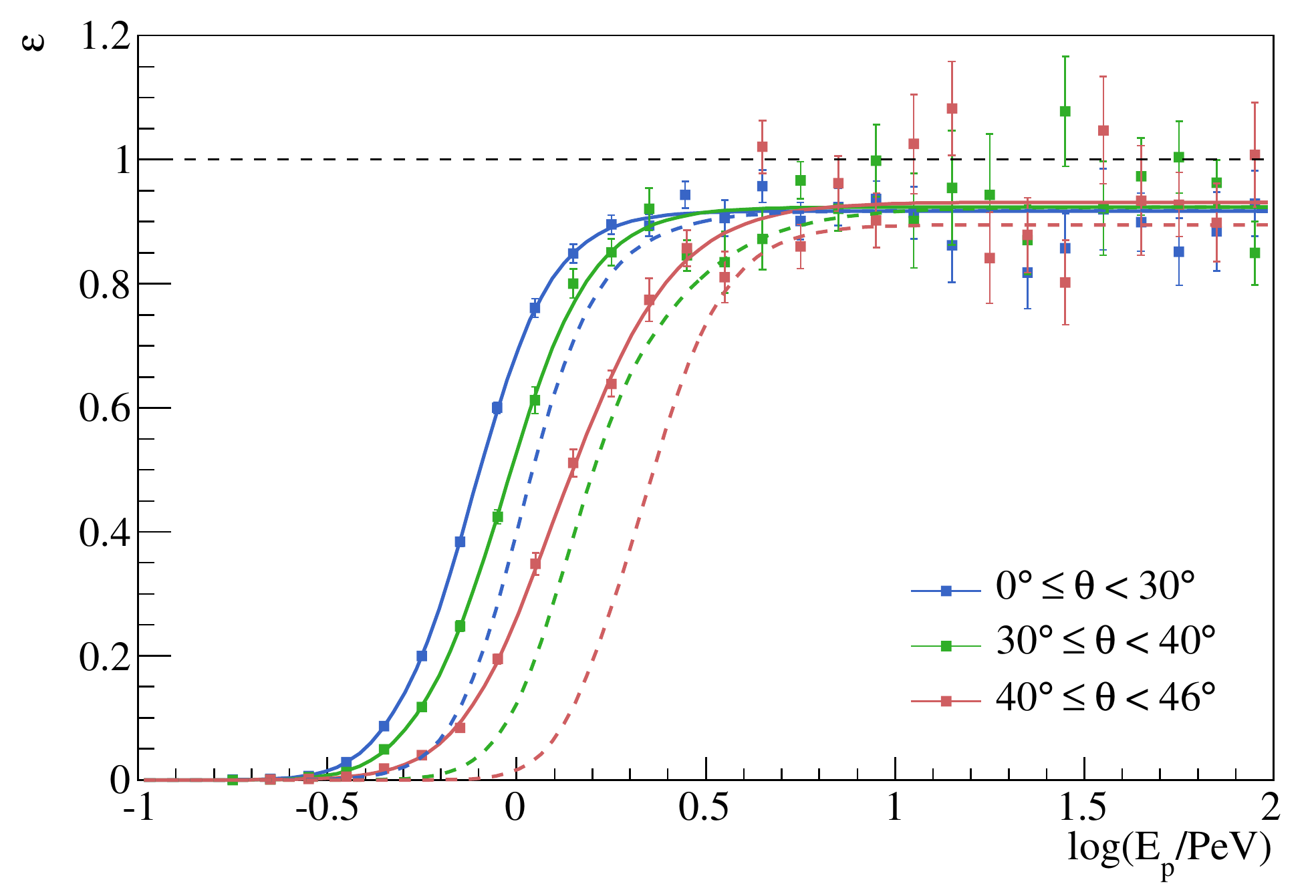}
  \caption{Total efficiency for proton showers in different zenith angle ranges as a function of energy. The solid lines are fits of function~\eqref{eq:efficiency_function}, and the dashed lines are the results of the corresponding fits for pure iron.}
  \label{fig:efficiency}
\end{figure}

\begin{figure}
  \centering
  \includegraphics[width=.6\columnwidth]{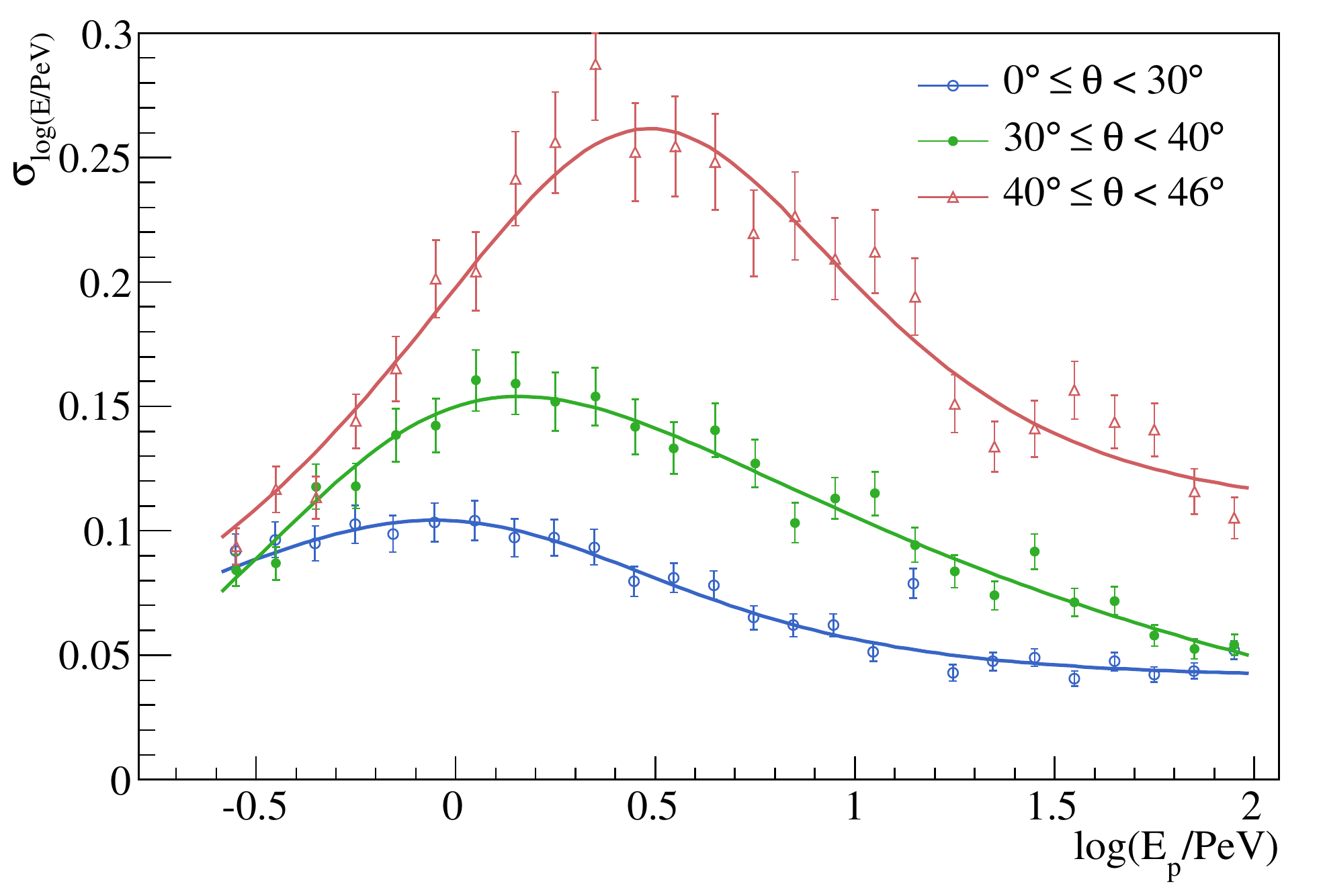}
  \caption{Energy resolution as defined in the text for proton showers in different zenith angle ranges. The lines are fits according to Eq.~\eqref{eq:sigma_function} in order to guide the eye.}
  \label{fig:resolution}
\end{figure}

Figure~\ref{fig:s125_vs_energy_proton} shows the response matrix for simulated proton and iron primaries in the interval $\Omega_1$ of smallest zenith angles. 
In each bin the colour code represents the probability that an event with energy~$E_p$ yields a signal~$S_{125}$. 
The binning uses a logarithmic scale. 

For computational purposes and to smooth fluctuations in the simulated response matrix, the $\log S_{125}$ projections of each~$\log E_p$ bin $j$ were fitted by a normal distribution function yielding the mean value $\langle\log S_{125}\rangle_j$ and standard deviation~$\sigma_{\log S, j}$. 
The normalisation~$\eps_j$
was calculated as the ratio between the sum of Monte Carlo event weights in the final sample and the sum of weights of events generated inside the fiducial area defined in Section~\ref{sub:final_selection}:
\begin{equation}\label{eq:efficiency}
  \eps_j = \frac{\sum_{i = 1}^{N_\mathrm{rec}}w_i}{\sum_{i = 1}^{N_\mathrm{gen}}w_i}.
\end{equation}
Due to migration of shower cores from outside the fiducial area, this quantity can become larger than unity. 
The energy dependence of the parameters $\langle\log S_{125}\rangle$, $\sigma_{\log S}$, and $\eps$ was then fitted by empirical functions (see~\ref{app:response_matrix}).
These functions were used to smooth statistical fluctuations in the response matrix and to extrapolate the range of simulations to higher energies in order to avoid potential artifacts that might be introduced by cutting the spectra off at $100\un{PeV}$.

The mean values and standard deviations are indicated in Fig.~\ref{fig:s125_vs_energy_proton} by the points with vertical bars. 
The response functions $\langle\log S_{125}(E_p)\rangle$ of proton showers for the three different zenith angular intervals $\Omega_k$ are shown in Fig.~\ref{fig:s125_vs_energy_proton_profiles}. 
Since the shower maximum lies above the detector throughout the covered energy range, showers from larger zenith angles are more strongly attenuated by the atmosphere and thus have a smaller shower size. 
In Fig.~\ref{fig:s125_vs_energy_p-fe}, these points are compared to the mean values for iron.
The response matrices for proton and iron are very different: on average, iron showers have their first interaction at larger height leading to a larger shower age than for protons. 
Iron showers yield a smaller average signal than proton showers with the same primary energy. 
The difference between proton and iron increases at larger zenith angles. 
This zenith angle dependence has been exploited to test the consistency of our data with models for the mass composition, as will be discussed in Section~\ref{sec:result}.

Figure~\ref{fig:efficiency} shows the efficiencies~$\eps$ obtained in the~$\log E_p$ bins, which are the normalisations of the normal distributions of $\log S_{125}$ belonging to this bin, for protons and iron nuclei comparing all zenith angle intervals. 
The lines are fits to Eq.~\eqref{eq:efficiency_function}. 
Mostly due to the very conservative containment criteria, peak efficiencies were significantly below $100\%$.
The maximum efficiencies in the three zenith angle ranges $\Omega_k$ correspond to the following effective areas:
\begin{align*}
  \Omega_1: A_{\rm eff} &= (1.051 \pm 0.013) \cdot 10^5\un{m^2} \\
  \Omega_2: A_{\rm eff} &= (0.900 \pm 0.019) \cdot 10^5\un{m^2} \\
  \Omega_3: A_{\rm eff} &= (0.803 \pm 0.012) \cdot 10^5\un{m^2}.
\end{align*}
Within statistical uncertainties the same values were obtained for iron primaries. 

The energy resolution (see Fig.~\ref{fig:resolution}) has been determined by transforming the~$\log S_{125}$ distribution for a given~$E_p$ back onto the $\log E_p$ axis.
It is worst where the detector becomes fully efficient, which happens between $1$ and $3\un{PeV}$ depending on zenith angle.
Towards higher energies the resolution improves, reaching values between $0.04$ and $0.12$ in $\log(E_p)$ at $100\un{PeV}$, corresponding to a resolution~$\sigma_E/E$ between~$9\%$ and~$23\%$.
The improvement of the energy resolution in the threshold region toward lower energies is a cutoff effect due to the fact that showers of those energies will only trigger the detector if they fluctuate upward. 
This resolution only covers the statistical fluctuations, systematic uncertainties are discussed later in Section~\ref{sec:systematics}.

The response matrices obtained by this method depend on the primary composition assumption, as well as the hadronic interaction models and the parametrization of the South Pole atmosphere assumed in the simulation.

\subsection{Unfolding} \label{sec:unfolding}
The response matrices were inverted using an iterative unfolding method based on Bayes' theorem described in Ref.~\cite{dagostini_unfolding}, which takes into account the total efficiency~$\eps$ and migration due to the fluctuations $\sigma_{\log(S)}$. 
Simply inverting the response matrix $\mathbf{R}$ would lead to unnatural fluctuations in the result.

Starting from a prior distribution $P_k(E_p^{(j)})$ in the $k$-th iteration, the inverse of the response matrix $\mathbf{R}^{-1}$ is constructed by inverting $P(S^{(i)}_{125}|E^{(j)}_p) = R_{ij}$ using Bayes' theorem:
\begin{equation}
  P_k(E^{(j)}_p|S^{(i)}_{125}) = \frac{P(S^{(i)}_{125}|E^{(j)}_p) \, P_k(E^{(j)}_p)}{\sum_\ell P(S^{(i)}_{125}|E^{(\ell)}_p) \, P_k(E^{(\ell)}_p)}.
\end{equation}
Then, an estimate of the energy spectrum $\hat{N}^e_{j,k}$ is obtained from the charge spectrum $N^s_i$:
\begin{equation}
  \hat{N}^e_{j,k} = \frac{1}{\eps_j} \sum_i N^s_i \, P_k(E^{(j)}_p|S^{(i)}_{125}).
\end{equation}
In the last step of the iteration, $P_k(E^{(j)}_p)$ is replaced by
\begin{equation}
  P_{k+1}(E^{(j)}_p) = \frac{\hat{N}^e_{j,k}}{\sum_\ell \hat{N}^e_{\ell,k}}.
\end{equation}
As initial prior, $P_0(E^{(j)}_p) \sim E_p^{-3}$ was chosen.

After each iteration, the unfolded spectrum was folded with the response matrix, $\tilde{N}^s_{i,k} = \sum_j R_{ij}N^e_{j,k}$, and compared to the measured shower size spectrum. 
A convergence criterion was then defined using the change in~$\chi^2$ between~$\tilde{N}^s_{i,k}$ and the measured shower size spectrum~$N^s$ between two iterations~$k$ and~\mbox{$k+1$}, as in~\cite{HolgerUlrichThesis}:
\begin{equation}
  \Delta \chi^2(k, k+1) = \chi^2(\tilde{N}^s_k, N^s) - \chi^2(\tilde{N}^s_{k+1}, N^s).
\end{equation}
This quantity decreases monotonically during the iteration process.
However, at $\Delta\chi^2(k, k+1) = 0$ the unfolding would be equivalent to simply inverting~$R_{i,j}^{(k)}$ and the unfolded spectrum would fluctuate unnaturally.
To avoid this, the iteration was terminated once $\Delta\chi^2(k, k+1)$ fell below a certain value~$\Delta\chi^2_\mathrm{term}$.
The value of this limit was determined beforehand using a simple toy simulation in which a known spectrum was folded with the response matrix and then, after adding statistical fluctuations, unfolded again. 
In every iteration step of this unfolding procedure, the unfolded spectrum was compared to the known true spectrum.
Finally, $\Delta\chi^2_\mathrm{term} = 1.1$ was chosen where the agreement with the true spectrum was best on average.

The error bars on the unfolded spectrum were determined by varying the shower size spectra within their statistical errors and repeating the unfolding. 
This was repeated $n=3000$ times and the statistical errors in bin $j$ were determined by comparing each unfolding result $N^{e(k)}$ to the average result $\langle N^e\rangle$:
\begin{equation}\label{eq:bin-errors}
  (\sigma^e_j)^2 = \frac{1}{n-1} \sum_{k=1}^n \bigl(N^{e(k)}_j - \langle N^e\rangle_j\bigr)^2.
\end{equation}
Similarly, bin-to-bin correlations were obtained:
\begin{equation}\label{eq:bin-correlations}
  \cov(i,j) = \frac{1}{n} \sum_{k=1}^n \bigl(N^{e(k)}_i - \langle N^e\rangle_i\bigr)\bigl(N^{e(k)}_j - \langle N^e\rangle_j\bigr).
\end{equation}

It was verified with a simple toy model that this algorithm correctly reproduces a true input spectrum, which was folded with the detector response and that the error determination is correct~\cite{FabianPhdThesis}.

\subsection{Correction for snow}\label{sec:energy:snow}
Snow accumulates constantly on top of the IceTop tanks, but a manual measurement of the snow height is only possible during the austral summer and therefore is done only once every year. 
The detector simulation took the snow depths measured in January, 2007, into account.
Data, on the other hand, were taken between June and October, 2007, when more snow had accumulated.
In order to estimate the effect of this difference, the detector response to proton showers with primary energies of $1\un{PeV}$, $10\un{PeV}$ and $30\un{PeV}$ and zenith angles $0\degrees$, $30\degrees$ and~$40\degrees$ was simulated assuming once the snow heights measured in January 2007 and once those measured in January 2008. 
In January 2007 the average snow depth on top of IceTop tanks was $20.5\un{cm}$, while in January 2008 the average height on top of the same tanks was $53.2\un{cm}$.
Assuming constant increase in snow depth and proportionality between $\log S_{125}$ and snow depth, shower sizes in August,~2007, were estimated. 
This lead to the following zenith angle dependent energy corrections relative to the simulations based on the January~2007 snow height measurement, which were applied to all unfolded energy spectra.
Within the statistical uncertainties, no energy dependence could be observed:
\begin{equation}\label{eq:systematics:energy-shift}\begin{split}
  \Omega_1:\quad&\Delta\log(E/\mathrm{PeV}) = 0.0368 \pm 0.0009,  \\
  \Omega_2:\quad&\Delta\log(E/\mathrm{PeV}) = 0.0440 \pm 0.0013,  \\
  \Omega_3:\quad&\Delta\log(E/\mathrm{PeV}) = 0.0513 \pm 0.0008.
\end{split}\end{equation}

\section{Systematic uncertainties} \label{sec:systematics}
All systematic errors are summarized in Table~\ref{tab:systematics}. 
In the following details about the determination of the uncertainties of the energy determination and the flux measurement will be given.

\subsection{Snow height}
To estimate the systematic error due to the energy correction for snow described in Section~\ref{sec:energy:snow}, snow accumulation was assumed proportional to wind speed.
The numbers obtained in this way were compared to those assuming constant growth of the snow depth (see above).
The result of this comparison was used as an estimate of the systematic error on energy determination due to snow height.

\subsection{Variations of the atmosphere} \label{sub:systematics:atm_variation}
As discussed in Section~\ref{sub:atmosphere}, variations of the atmosphere affect the observed shower sizes.
The influence of two parameters of the atmosphere has been studied in a data driven way: the total overburden $X_0$, and the altitude profile~\mbox{$\dd X_v(h)/\dd h$}.

First, the days of data taking were ordered according to the total atmospheric overburden~$X_0$.
Then the 50 days with the highest and the 50 days with the lowest overburden were selected from the total of 153 days. 
The average overburdens during these periods were $X_\mathrm{low} = 679\un{g/cm^2}$ and $X_\mathrm{high} = 700\un{g/cm^2}$, yielding a difference of $\Delta X = 21\un{g/cm^2}$. 
From the data taken during these days shower size spectra were created for each zenith range $\Omega_k$. 

By comparing the shower size spectra obtained in the two periods, the dependence of $S_{125}$ with atmospheric overburden was derived.
The RMS variation of the total atmospheric overburden between June~1 and October~31, 2007, of~$\sigma_{X_v} = 9.86\un{g/cm^2}$ was used to estimate the systematic error on the energy determination due to atmospheric overburden variations.
With the given statistical precision, an energy dependence of this variation could not be observed.

In contrast to total overburden the altitude profile of the atmosphere at South Pole undergoes a clear annual cycle.
To study the effect of varying the atmospheric profile on air shower measurements the data taking period was divided into a period of very dense atmosphere (July~25th to October~10th) and one when the atmosphere was less dense (remaining days between June~1st and July~24th and between October~11th and October~31st).
Shower size spectra were extracted from the data taken in these two periods and by comparing those spectra, an additional systematic error due to the atmospheric profile variation was derived.

\subsection{Atmosphere model in simulation}
The CORSIKA simulations used a model of the South Pole atmosphere.
A systematic uncertainty arises from the choice of model since it does not exactly match the average atmosphere during the data taking period.
To estimate this error on the energy scale simulations two different atmosphere parametrizations were compared.
CORSIKA atmosphere model~12 (July~1, 1997), which was used in the unfolding procedure, has a total overburden of $692.9\un{g/cm^2}$ and atmosphere model~13 (October~1, 1997) has a total overburden of $704.4\un{g/cm^2}$.
Averaging the difference in~$\log S_{125}$ between the two simulations above~$E_p = 1\un{PeV}$ the systematic error due to the difference of the simulated overburden and the average overburden in data was determined.

\subsection{Calibration}\label{sub:systematics-vem}
Systematic uncertainties due to calibration can arise for two reasons: variations of the calibration constants between calibration runs, and a discrepancy between the calibration of the experiment and the detector simulation. 

\begin{figure}
  \centering
  \includegraphics[width=.6\columnwidth]{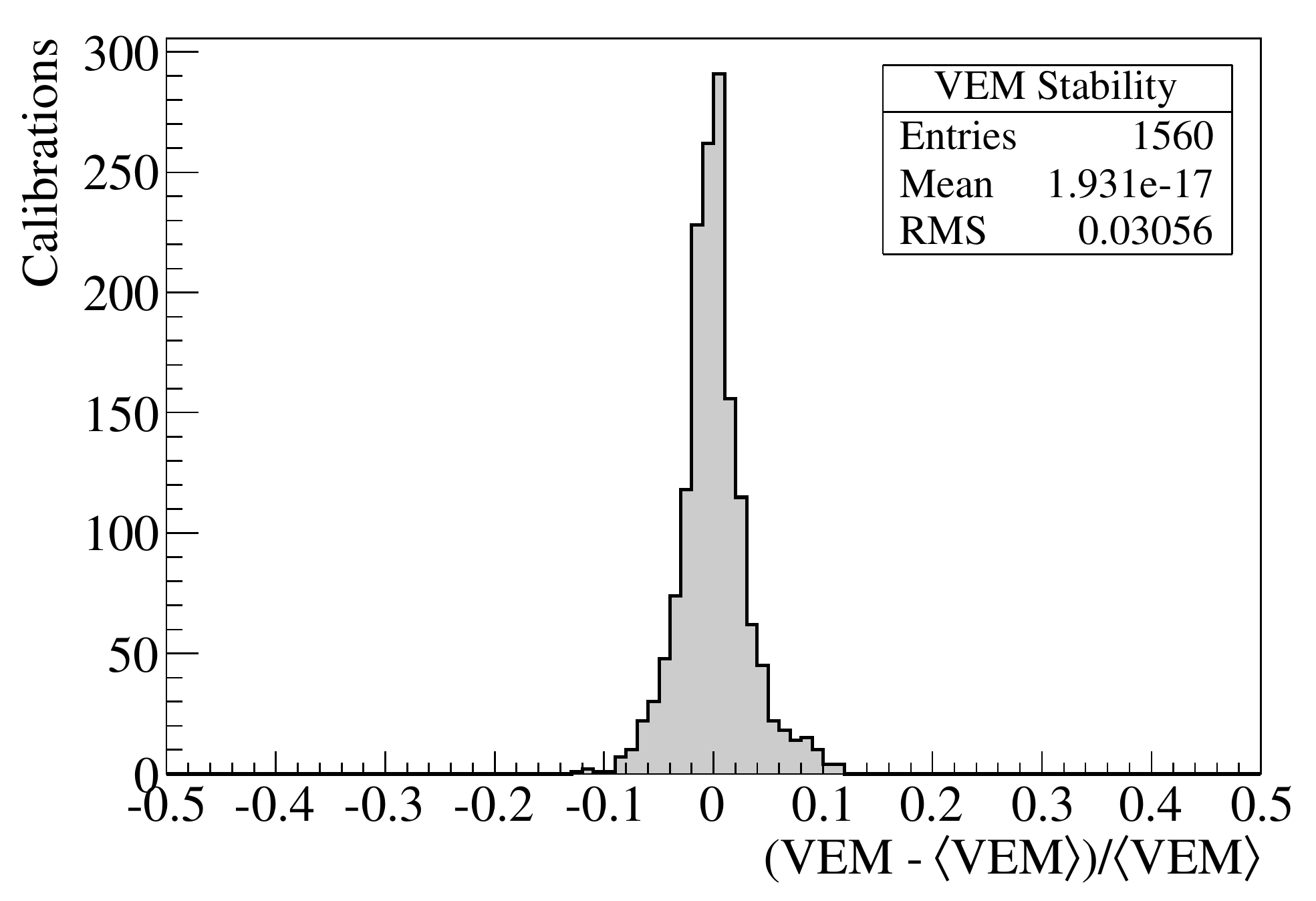}
  \caption{Relative change in the number of photoelectrons corresponding to $1\un{VEM}$ between consecutive calibration runs for all 15 calibration runs and all high and low-gain DOMs. Calibration runs were carried out every two weeks. The RMS value of 0.031 of this distribution was used to estimate the systematic error on energy determination.}
  \label{fig:vem_variation}
\end{figure}

The first point was addressed by studying the variation of the VEM calibration between calibration runs. 
Figure~\ref{fig:vem_variation} shows the relative difference in number of photoelectrons corresponding to~$1\un{VEM}$ between calibration runs for all DOMs.
From the RMS of this distribution the systematic uncertainty on the energy reconstruction due to variations of the VEM calibration was estimated to be $3.0\%$.

The simulated tanks were calibrated using the same procedure as for the real tanks, as described in Section~\ref{sub:detsim}. 
The conversion factor between Cherenkov photons and photo electrons resulting from this calibration has a statistical uncertainty of~$1.5\%$, which was included as a systematic error on the energy.

\subsection{Droop}\label{sub:systematics-droop}
The toroid used to decouple the PMT from the signal capture electronics introduces a significant droop effect (see Section~\ref{sub:calibration}), which was not corrected for in the analysis.
Not correcting for droop is not a source of systematic uncertainty in itself if it is done consistently in data and simulation. 
However, discrepancies in the way the droop effect is simulated in the detector Monte Carlo, may lead to undesired systematic effects. 
In order to quantify these effects, the effect of a droop correction algorithm on the recorded charges was compared between data and simulation.
From this comparison a systematic error on the energy determinatino of~$1.5\%$ was derived.

\subsection{PMT Saturation}
Inaccuracies in the simulated saturation behaviour of the PMT could introduce systematic uncertainties on the energy determination mostly at high energies.
In simulation, saturation sets in at higher charges than in the experiment.
In order to estimate the effect of this discrepancy on the energy spectrum, an artificial, charge-based saturation function was applied to the simulated charges to bring the simulated charge spectrum into agreement with experimental data.
Then, the simulated showers were reprocessed, and the change in $\log(S_{125}/\mathrm{VEM})$ was used to estimate the systematic error on the energy.
For primary energies below $10\un{PeV}$, the systematic error due to the difference in saturation behaviour is less than $0.5\%$. 
Above $10\un{PeV}$ it increases exponentially to a value of $2.5\%$ at $100\un{PeV}$.

\subsection{Cut efficiencies}\label{sub:systematics-cuts}
Differences in the effects of quality cuts described in Section~\ref{sub:final_selection} when applied to experimental and simulated data lead to a systematic uncertainty on the efficiency and consequently on the flux normalization.
Passing rates of all cuts for data and Monte Carlo events above threshold are listed in Table~\ref{tab:cutflow}.
There is a relative difference of $3.0\%$ between data and Monte Carlo in the total cumulative passing rate, which is included in the systematic uncertainty on the flux.

\subsection{Interaction model}

\begin{figure}
  \centering
  \includegraphics[width=.6\columnwidth]{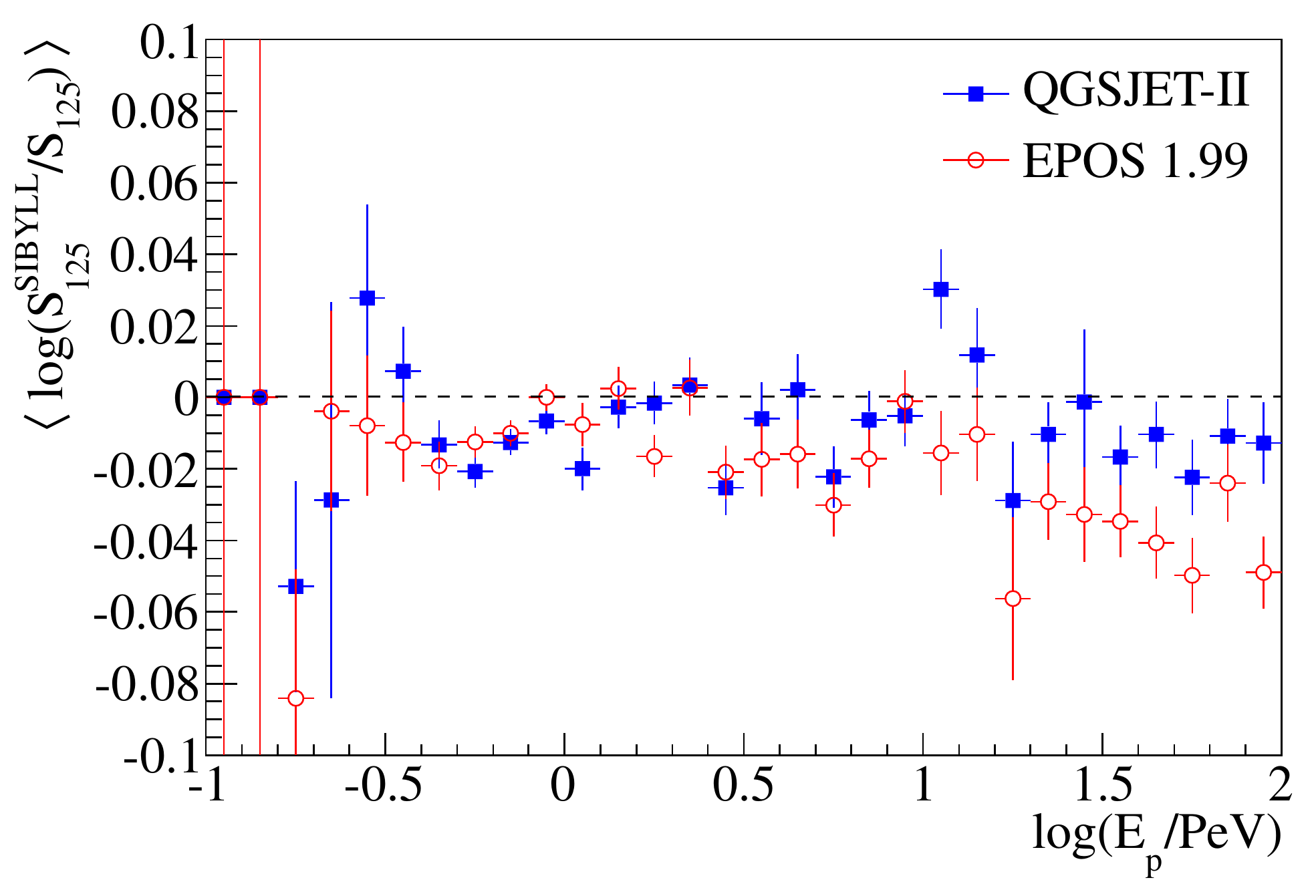}
  \caption[Shower size ratio for the two-component assumption between SIBYLL and QGSJET and EPOS based simulations]
    {Shower size ratio for the two-component assumption between SIBYLL and QGSJET and EPOS based Monte Carlo simulations for the two component primary assumption and showers with zenith angles up to $30\degrees$.}
  \label{fig:systematics:hadronic-model}
\end{figure}

Small simulation datasets of proton and iron showers created using the high energy hadronic interaction models QGSJET-II and EPOS~1.99 in addition to SIBYLL were used to estimate the systematic uncertainty due to the modeling of hadronic interactions. 
Figure~\ref{fig:systematics:hadronic-model} shows the shower size ratio between SIBYLL and the alternative simulations as a function of primary energy for the two-component primary composition assumption and zenith angles up to~$30\degrees$.
Simulations with SIBYLL seem to yield systematically smaller shower sizes, and the same observation was made for more inclined showers.

The systematic error derived in this way is purely based on a comparison of the three interaction models.
All of these models have different known strengths and weaknesses in their description of the underlying physics.
Additionally, they all include extrapolations of cross-sections and multiplicity distributions to energy ranges not accessible by current collider experiments which are relevant in the first few cosmic ray interactions.
Thus, there is an unknown systematic error in case the range hadronization models does not cover the true behavior.

\subsection{Response matrix}\label{sub:systematics-response}
Limited Monte Carlo statistics introduce uncertainties into the response matrix. 
Assuming the efficiency is constant above the threshold, the flux error induced by uncertainties of the detector response can be estimated by the fit error on $c_0$ in Eq.~\eqref{eq:efficiency_function}.
The uncertainties on the parameters $a_0$ and $b_0$ in Equations~\eqref{eq:s125_function} and~\eqref{eq:sigma_function} translate to an uncertainty on the energy in the unfolding process.
These statistical uncertainties on the response matrix were also included in the systematic error of the final result.

Additionally, the flux model used in the simulation also influences the response matrix. 
A harder spectrum leads to larger average shower sizes in an energy bin than a softer one.
Simulations based on an~$E^{-2}$~flux and an~$E^{-4}$~flux were compared with the standard simulation which assumes a power law of $E^{-3}$.
Above the threshold the resulting difference in shower size appears to be independent of primary energy.
The differences in shower size between the two extreme spectral indices were used as an estimate of the systematic error on energy scale due to the assumed flux model.

\subsection{Unfolding procedure}\label{sub:systematics:unfolding}
Two parameters besides the response matrix influence the result of the unfolding: the termination criterion $\Delta\chi^2_\mathrm{max}$ and the prior distribution $P_0$.
Varying the termination criterion, lead to a variation of the total flux, which was included as a systematic error.

In addition, varying the spectral index of the initial prior~$P_0$ between~$-2.5$ and~$-3.5$, a variation of the total flux of about~$2\%$ was observed. 
Below the knee region around~$3$ to~$4\un{PeV}$, the spectral index seems to depend on the prior (in the most inclined zenith interval even up to~$10\un{PeV}$.
Varying the prior lead to a variation of the spectral index below the knee in the most vertical zenith band by~${\pm}0.01$, and in the most inclined zenith range by~${\pm}0.025$.
At higher energies variations appear to be purely statistical.

\subsection{Summary of systematic errors}
Systematic uncertainties are summarized in Table~\ref{tab:systematics}.
The total systematic uncertainty was determined by quadratically adding the individual contributions.
The error on the determination of the primary energy in the most vertical zenith angle range is~$5.1\%$ below $E_p=10\un{PeV}$, and~$5.7\%$ above. 
Main contributions are the calibration stability~($3.0\%$), atmosphere~($2.7\%$ in total), and the hadronic interaction model~($2.1\%$).
The systematic influence of the unknown primary composition will be discussed in the next section.
Furthermore, a flux uncertainty of~$3.5\%$ is caused by differences in cut efficiencies between data and Monte Carlo, the efficiency calculation in Monte Carlo, and the termination criterion and seed in the unfolding procedure.

\begin{table*}
  \caption{Summary of systematic uncertainties of the energy and flux determination in the three zenith angle intervals $\Omega_1$, $\Omega_2$, and $\Omega_3$. The individual points are explained in the text.}
  \vspace*{1ex}
  \label{tab:systematics}
  \centering
  \begin{tabular}{lr@{.}lr@{.}lr@{.}lr@{.}lr@{.}lr@{.}l}
    \toprule
      & \multicolumn{4}{c}{$\boldsymbol{0\degrees \leq \theta < 30\degrees}$}
        & \multicolumn{4}{c}{$\boldsymbol{30\degrees \leq \theta < 40\degrees}$}
        & \multicolumn{4}{c}{$\boldsymbol{40\degrees \leq \theta < 46\degrees}$} \\
    \textbf{Uncertainty}
        & \multicolumn{2}{c}{\textbf{Energy}} & \multicolumn{2}{c}{\textbf{Flux}}
        & \multicolumn{2}{c}{\textbf{Energy}} & \multicolumn{2}{c}{\textbf{Flux}}
        & \multicolumn{2}{c}{\textbf{Energy}} & \multicolumn{2}{c}{\textbf{Flux}} \\
    \midrule
    Snow height              &  0 & 4\% & \multicolumn{2}{l}{}
                             &  0 & 4\% & \multicolumn{2}{l}{}
                             &  0 & 4\% & \multicolumn{2}{l}{} \\
    Overburden variation     &  0 & 26\% & \multicolumn{2}{l}{}
                             &  1 & 9\% & \multicolumn{2}{l}{}
                             &  3 & 0\% & \multicolumn{2}{l}{} \\
    Atmosphere profile variation &  2 & 5\% & \multicolumn{2}{l}{}
                             &  1 & 8\% & \multicolumn{2}{l}{}
                             &  1 & 1\% & \multicolumn{2}{l}{} \\
    Atmosphere model         &  0 & 9\% & \multicolumn{2}{l}{}
                             &  1 & 1\% & \multicolumn{2}{l}{}
                             &  0 & 6\% & \multicolumn{2}{l}{} \\
    MC Calibration           &  1 & 5\% & \multicolumn{2}{l}{}
                             &  1 & 5\% & \multicolumn{2}{l}{}
                             &  1 & 5\% & \multicolumn{2}{l}{} \\
    PMT saturation, $E_p \leq 10\un{PeV}$
                             &  0 & 5\% & \multicolumn{2}{l}{}
                             &  0 & 5\% & \multicolumn{2}{l}{}
                             &  0 & 5\% & \multicolumn{2}{l}{} \\
    PMT saturation, $E_p > 10\un{PeV}$
                             &  $<$2 & 5\% & \multicolumn{2}{l}{}
                             &  $<$2 & 5\% & \multicolumn{2}{l}{}
                             &  $<$2 & 5\% & \multicolumn{2}{l}{} \\
    Droop                    &  1 & 5\% & \multicolumn{2}{l}{}
                             &  1 & 5\% & \multicolumn{2}{l}{}
                             &  1 & 5\% & \multicolumn{2}{l}{} \\
    Calibration stability    &  3 & 0\% & \multicolumn{2}{l}{}
                             &  3 & 0\% & \multicolumn{2}{l}{}
                             &  3 & 0\% & \multicolumn{2}{l}{} \\
    Interaction model        &  2 & 1\% & \multicolumn{2}{l}{}
                             &  4 & 3\% & \multicolumn{2}{l}{}
                             &  2 & 0\% & \multicolumn{2}{l}{} \\
    Flux model               & 0 & 7\% & \multicolumn{2}{l}{}
                             & 1 & 0\% & \multicolumn{2}{l}{}
                             & 1 & 0\% & \multicolumn{2}{l}{} \\
    $\langle \log S \rangle$ and $\sigma_{\log S125}$
                             & 0 & 7\% & \multicolumn{2}{l}{}
                             & 1 & 2\% & \multicolumn{2}{l}{}
                             & 0 & 8\% & \multicolumn{2}{l}{} \\
    \midrule
    Cut passing rates        & \multicolumn{2}{l}{} & 3 & 0\%
                             & \multicolumn{2}{l}{} & 3 & 0\%
                             & \multicolumn{2}{l}{} & 3 & 0\% \\
    Efficiency               & \multicolumn{2}{l}{} & 0 & 9\%
                             & \multicolumn{2}{l}{} & 1 & 6\%
                             & \multicolumn{2}{l}{} & 1 & 2\% \\
    Unfolding procedure      & \multicolumn{2}{l}{} & 1 & 6\%
                             & \multicolumn{2}{l}{} & 3 & 4\%
                             & \multicolumn{2}{l}{} & 5 & 2\% \\
    \midrule
    \textbf{Total:} $\boldsymbol{E_p \leq 10}\,\mathbf{PeV}$ 
                             & \multicolumn{1}{r@{\bf .}}{\bf 5} & {\bf 1\%}
                               & \multicolumn{1}{r@{\bf .}}{\bf 3} & {\bf 5\%}
                             & \multicolumn{1}{r@{\bf .}}{\bf 6} & {\bf 5\%}
                               & \multicolumn{1}{r@{\bf .}}{\bf 4} & {\bf 8\%}
                             & \multicolumn{1}{r@{\bf .}}{\bf 5} & {\bf 5\%}
                               & \multicolumn{1}{r@{\bf .}}{\bf 6} & {\bf 1\%} \\
    \phantom{\textbf{Total:} }$\boldsymbol{E_p > 10}\,\mathbf{PeV}$
                             & \multicolumn{1}{r@{\bf .}}{\bf 5} & {\bf 7\%}
                               & \multicolumn{1}{r@{\bf .}}{\bf 3} & {\bf 5\%}
                             & \multicolumn{1}{r@{\bf .}}{\bf 7} & {\bf 0\%}
                               & \multicolumn{1}{r@{\bf .}}{\bf 4} & {\bf 8\%}
                             & \multicolumn{1}{r@{\bf .}}{\bf 6} & {\bf 0\%}
                               & \multicolumn{1}{r@{\bf .}}{\bf 6} & {\bf 1\%} \\
    \bottomrule
  \end{tabular}
\end{table*}

\section{Energy spectrum}\label{sec:result}

\begin{figure}
  \centering
  \subfigcapskip-3ex
  \subfigbottomskip3ex
  \subfigure[Proton assumption]{
    \includegraphics[width=.6\columnwidth]{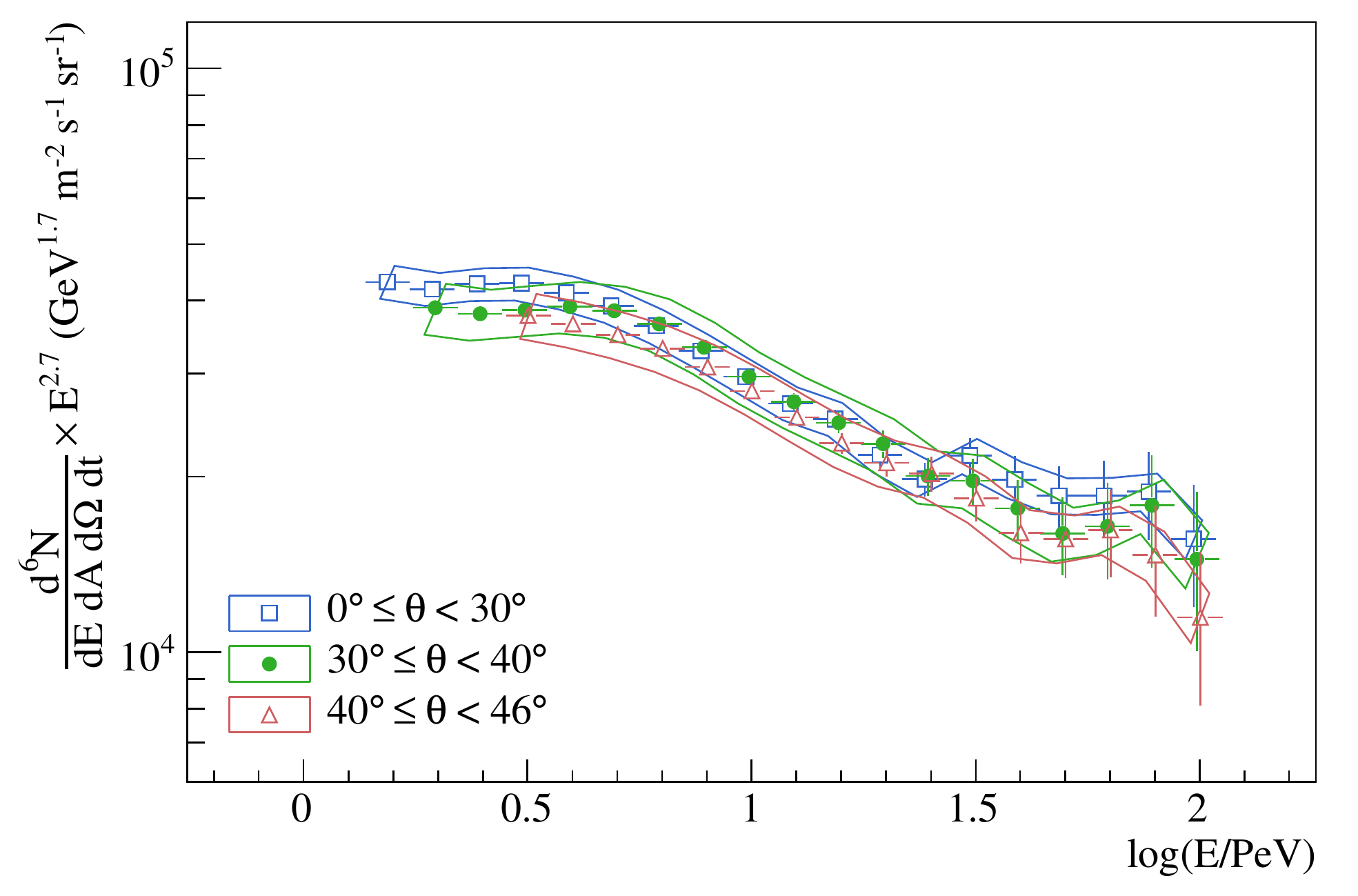}
    \label{fig:proton_result}}
  \subfigure[Iron assumption]{
    \includegraphics[width=.6\columnwidth]{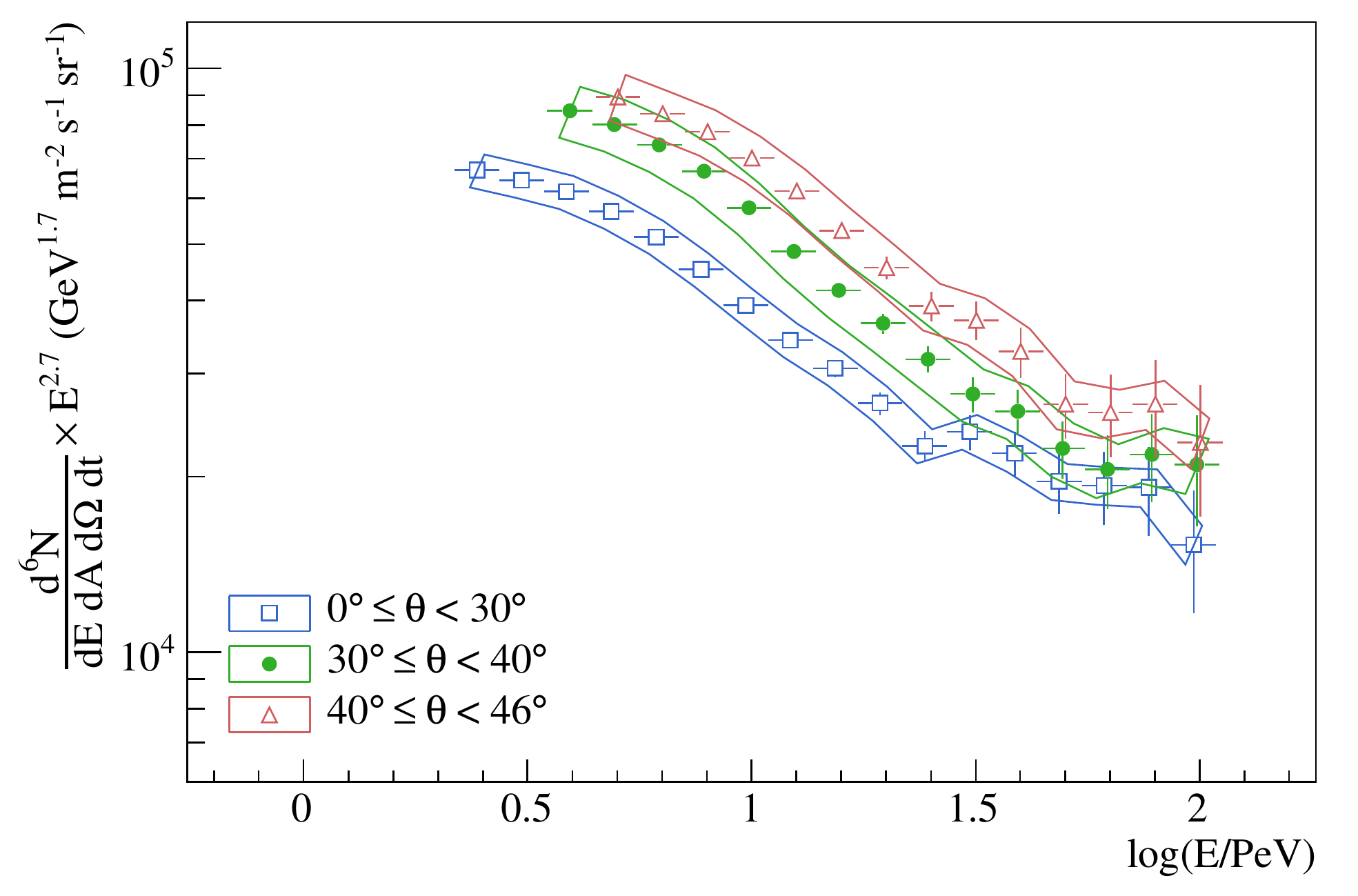}
    \label{fig:iron_result}}
  \subfigure[Two-component model]{
    \includegraphics[width=.6\columnwidth]{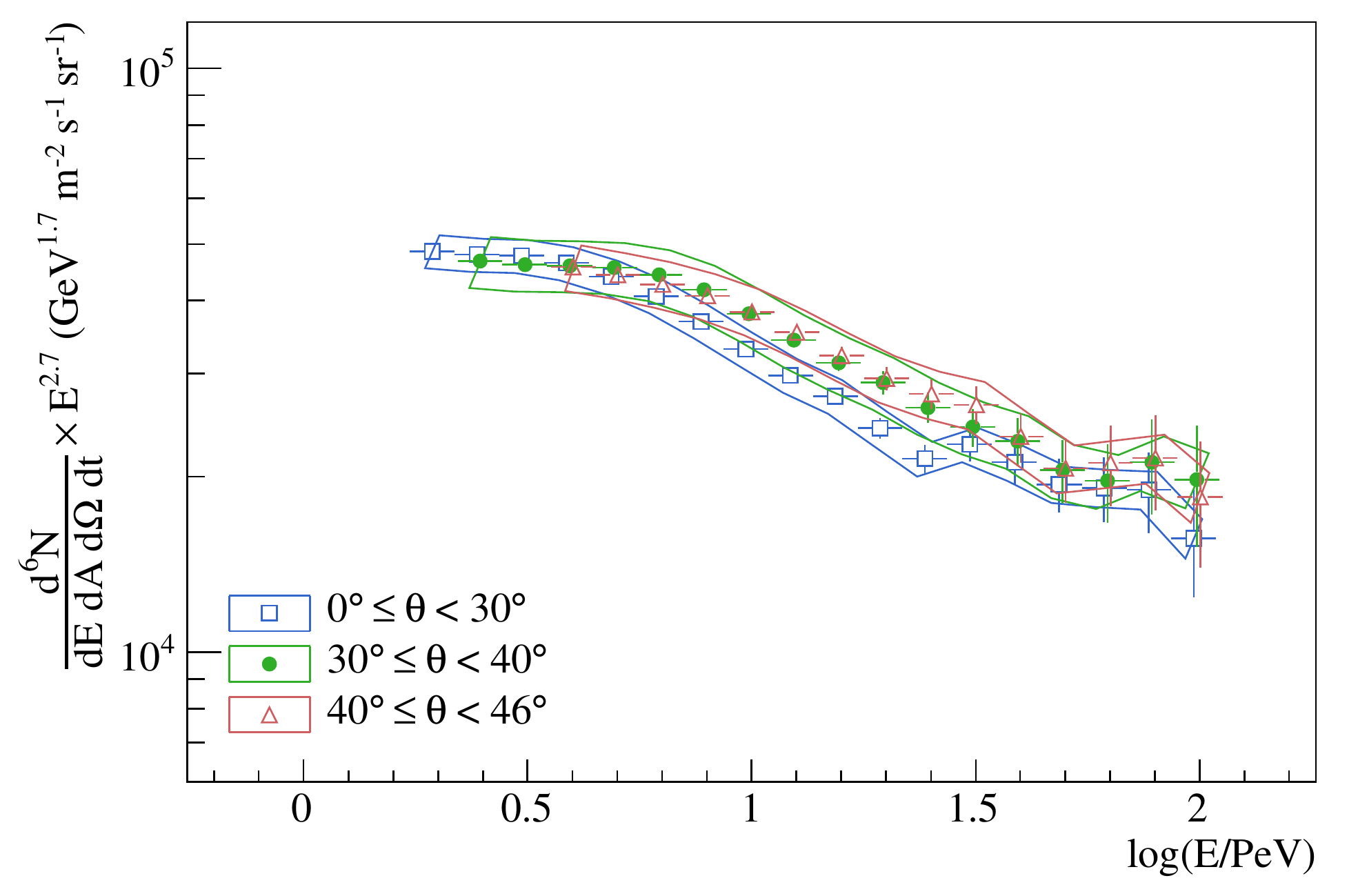}
    \label{fig:twocomponent_result}}
  \caption{Resulting flux measured with IceTop, weighted with $E^{1.7}$. The reconstruction was done using three different composition assumptions as described in the text: (a)~pure proton, (b)~pure iron, and (c)~Glasstetter's two-component model. In each case, the data were divided into three different zenith angle bands equidistant in~$\sec(\theta)$. Based on the assumption of an isotropic flux, the three individual spectra should agree. The boxes indicate the systematic errors.}
  \label{fig:result_spectra}
\end{figure}

Figure \ref{fig:result_spectra} shows energy spectra for three zenith angular intervals unfolded under three assumptions on the mass composition: all-proton, all-iron and the two-component model~\cite{glasstetter} explained in Section~\ref{sub:simulation_datasets}. 
The lower end of the energy range of each spectrum was selected where the efficiency according to Eq.~\eqref{eq:efficiency_function} reached $90\%$ of the maximum value. 
The threshold was determined individually for each zenith interval and primary composition assumption.
That way the threshold region is excluded and the efficiency can be assumed almost constant. 
Based on the energy resolution, a binning of 10 bins per decade was chosen.

In Fig.~\ref{fig:s125_vs_energy_p-fe} it was shown that the difference in shower size between simulated proton and iron showers increases with zenith angle due to the increasing slant depth in the atmosphere, which has a different effect for the different masses: iron showers are attenuated more strongly with increasing slant depth than proton showers. 
Since the cosmic-ray flux is isotropic to a few per thousand the flux measured in different zenith angular intervals has to be the same.

In case of the pure proton assumption (Fig.~\ref{fig:proton_result}) a good agreement between the three spectra is observed.
Assuming pure iron (Fig.~\ref{fig:iron_result}), the individual spectra for the three different zenith bands clearly disagree at low energies while they start to converge toward higher energies.
Agreement of the three spectra in case of the two-component model (Fig.~\ref{fig:twocomponent_result}) is good at low and high energies.
In the intermediate energy range there is some deviation between the spectrum obtained from steepest zenith angle range and the other two spectra.
However, they are still consistent when considering systematic uncertainties.

\begin{figure}
  \centering
  \includegraphics[width=.6\columnwidth]{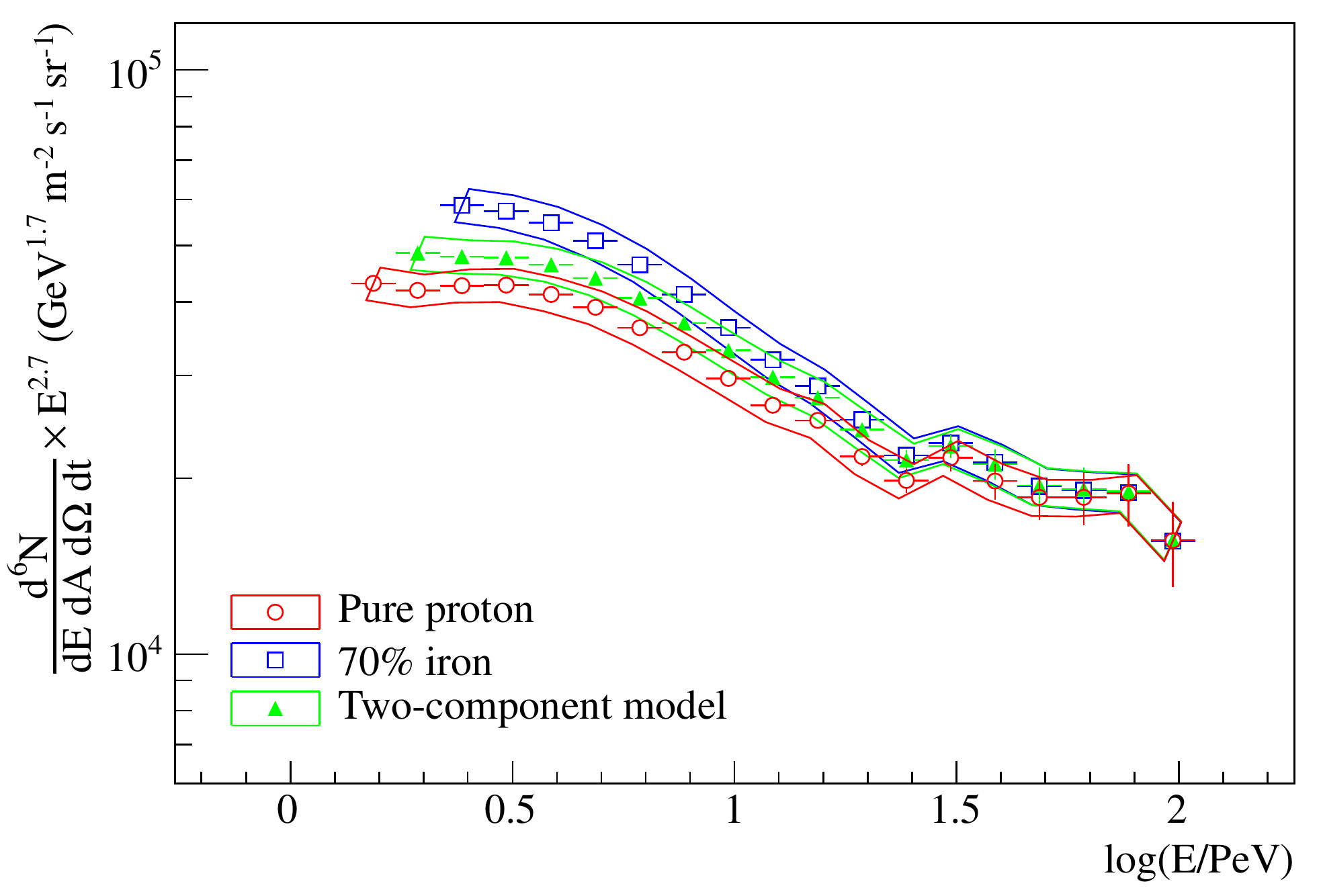}
  \caption{All particle spectra obtained with IceTop from air showers with zenith angles up to~$30^\circ$ under three different composition assumptions: pure proton, the two-component model, and a mixture of~$30\%$ proton and~$70\%$ iron.}
  \label{fig:modelcomparison_result}
\end{figure}

Using a~$\chi^2$~comparison of fluxes in each bin of the spectra from the three zenith angle ranges, pure iron could be excluded at a~${>}99\%$ confidence level below~$25\un{PeV}$.
This comparison took into account both statistical and systemtic errors.
The latter were treated in a conservative way by assuming no correlations between them for the different zenith angle intervals.
Using the same comparison and various mixtures of proton and iron, up to~$70\%$ of iron cannot be excluded at any energy.

In Fig.~\ref{fig:modelcomparison_result}, the results obtained in the steepest zenith angle range~$\Omega_1$ with three primary composition assumptions are compared: pure proton, the two-component model, and~$70\%$~iron.
Only the most vertical zenith angle range was chosen, because the difference in size for showers initiated by different primaries is smallest in this zenith interval, as seen in Fig.~\ref{fig:s125_vs_energy_p-fe}, and because systematic uncertainties are smallest in this range.
Because the difference in shower size between proton and iron decreases toward higher energies, the spectrum obtained under the~$70\%$~iron assumption is softer than the proton-based result.
While the composition model has a sizable influence on the measured all-particle flux below~$10\un{PeV}$, the difference between the two extreme assumptions of pure proton and~$70\%$~iron almost disappears above~$30\un{PeV}$.

As a final result the cosmic ray spectrum is given separately for the assumptions of the pure-proton and the two-component model which both yield consistent fluxes in the different zenith angle ranges. 
The systematic errors, as depicted in Fig.~\ref{fig:modelcomparison_result} by the bands covering the data points, are evaluated for all assumptions separately and without including the uncertainty from the unknown composition.
The~$70\%$-iron case was used in addition to determine the systematic error range on the flux due to primary composition. 
The range of systematic errors lies between the upper border of the~$70\%$-iron band and the lower border of the pure-proton band. 
At~$2.4\un{PeV}$, for example, the allowed fluxes range from~$2.65 \times 10^{-13}$ to~$3.34 \times 10^{-13}\un{GeV^{-1} \, m^{-2} \, s^{-1} \, sr^{-1}}$. 
The contribution to the systematic uncertainty due to the primary composition decreases from about~$30\%$ at~$2\un{PeV}$ to less than~$1\%$ above about~$60\un{PeV}$.

\begin{figure*}
  \centering
  \includegraphics[width=0.8\textwidth]{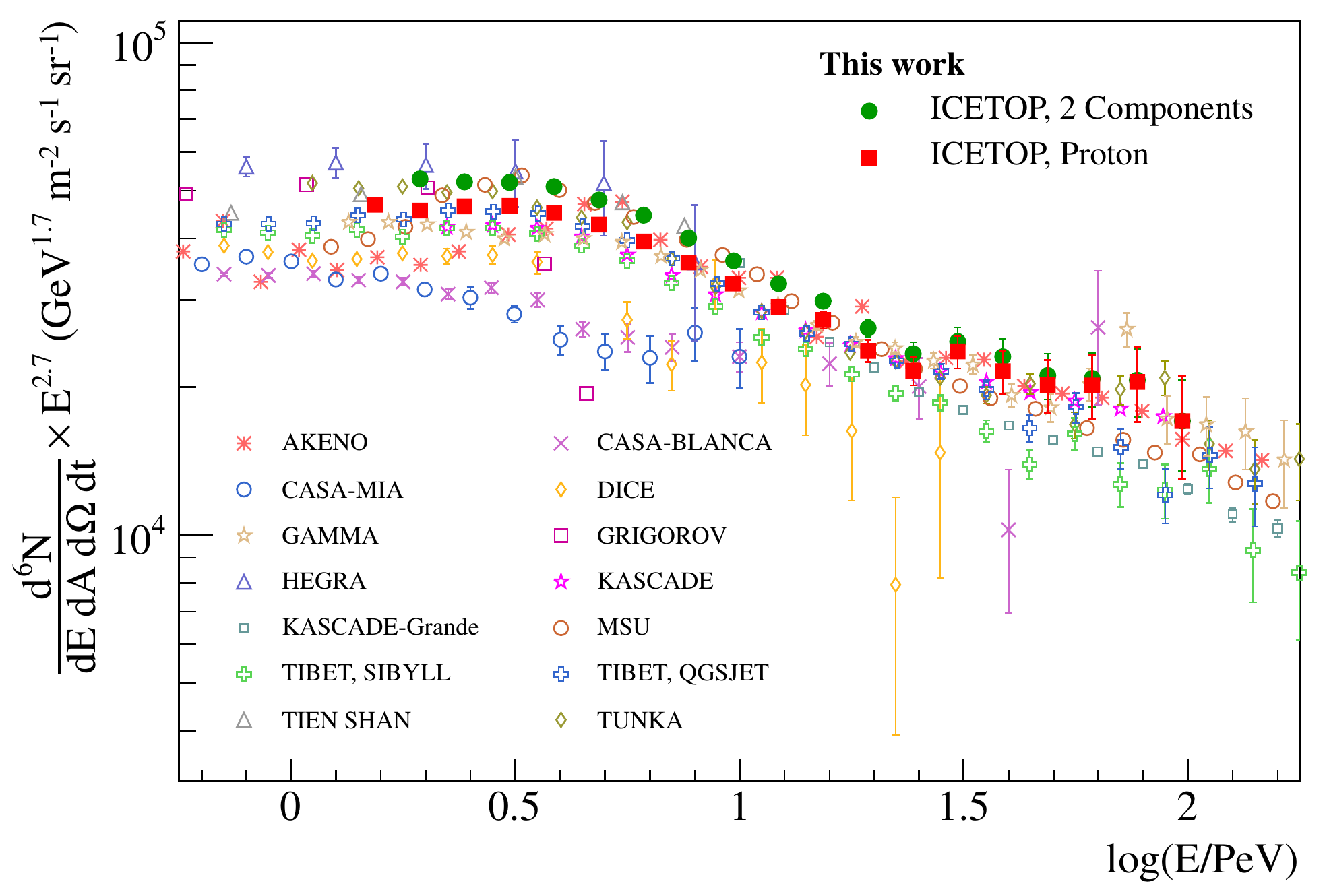}
  \caption{The all-particle cosmic ray energy spectrum obtained from the analysis of IceTop data of events with zenith angles up to $30\degrees$ compared to a selection of other experimental results~\cite{Nagano84,Nagano92,Fowler01,Glasmacher99,SwordyKieda00,Garyaka08,Grigorov70,Grigorov71,Arqueros00,KascadeSpectrum,Haungs09,Fomin96,Amenomori08,Antonov95,Antokhonov10}.}
  \label{fig:final_result}
\end{figure*}

Figure~\ref{fig:final_result} shows the results for pure proton and the two-component model, without the systematic error bands, in comparison to a selection of other experiments~\cite{Nagano84,Nagano92,Fowler01,Glasmacher99,SwordyKieda00,Garyaka08,Grigorov70,Grigorov71,Arqueros00,KascadeSpectrum,Haungs09,Fomin96,Amenomori08,Antonov95,Antokhonov10}.
Table~\ref{tab:result} lists the measured fluxes for these two primary composition assumptions.
The systematic errors on the flux given in the table have been calculated by transforming the systematic error on energy into a flux error based on the local spectral index $\gamma$: $\Delta I/I = \gamma \Delta E/E$.
This was added quadratically to the systematic error on the flux.

The two spectra have been fitted with the following parametrization~\cite{TerAntonyanFlux}:
\begin{equation}\label{eq:spectrum:hoerandel}
  \frac{\dd N}{\dd\ln E} = \frac{I_\mathrm{knee}}{2^{(\gamma_2-\gamma_1)/\eps}} \, \left(\frac{E}{E_\mathrm{knee}}\right)^{\gamma_1+1} \left(1 + \left(\frac{E}{E_\mathrm{knee}}\right)^\eps\right)^{(\gamma_2-\gamma_1)/\eps},
\end{equation}
where $I_\mathrm{knee}$ is the flux at the knee, $E_\mathrm{knee}$ is the position of the knee, $\gamma_1$ is the spectral index below and $\gamma_2$ above the knee, and~$\eps$ describes the sharpness of the knee.
In the fit, statistical errors and bin-to-bin correlations according to Equations~\eqref{eq:bin-errors} and~\eqref{eq:bin-correlations} were used.
The results are listed in Table~\ref{tab:fitparams}.

In pure-proton case, the data points below the knee are not well fitted with the assumption of a single slope.
This could either be a real feature of the spectrum or an indication of a wrong composition because in the region of the first two points the
energy threshold causes a mass dependent efficiency. 
In order to obtain nevertheless also for the pure-proton case a fit with a unique slope below the knee, the first two data points have been excluded from the fit.
The variation of the parameters when including the first or the second point respectively was used as a systematic error. 
When including all data points, all parameters lie within this range, but only a bad fit is achieved.

\begin{table*}
  \centering
  \caption{All-particle cosmic ray energy spectra measured by the IceTop air shower array for the pure proton and the two-component primary composition assumptions using the hadronic interaction model SIBYLL2.1.}
  \vspace*{2ex}
  \begin{tabular}{r@{.}lr@{.}l@{$\,\pm\,$}r@{.}l@{$\,\pm\,$}r@{.}l@{$\,\times\,$}lr@{.}l@{$\,\pm\,$}r@{.}l@{$\,\pm\,$}r@{.}l@{$\,\times\,$}l}
    \toprule
    \multicolumn{2}{c}{\textbf{Energy}} & 
        \multicolumn{14}{c}{$\boldsymbol{\dd N/\dd E \pm \text{\textbf{stat}} \pm \text{\textbf{syst}}}$ 
                     \textbf{($\mathbf{GeV^{-1} \, m^{-2} \, s^{-1} \, sr^{-1}}$)}} \\
    \cmidrule{3-16}
    \multicolumn{2}{c}{\textbf{($\mathbf{10^6\,GeV}$)}} & \multicolumn{7}{c}{\textbf{Proton}} & \multicolumn{7}{c}{\textbf{Two-component}} \\
    \midrule
     $1$ & $54$ & $(9$ & $26$  & $0$ & $04$  & $1$ & $4)$  & $10^{-13}$ & \multicolumn{7}{c}{} \\
     $1$ & $94$ & $(4$ & $838$ & $0$ & $020$ & $0$ & $7)$  & $10^{-13}$ & $(5$ & $612$  & $0$ & $018$ & $0$ & $9)$  & $10^{-13}$ \\
     $2$ & $44$ & $(2$ & $650$ & $0$ & $014$ & $0$ & $4)$  & $10^{-13}$ & $(2$ & $974$  & $0$ & $012$ & $0$ & $5)$  & $10^{-13}$ \\
     $3$ & $07$ & $(1$ & $426$ & $0$ & $010$ & $0$ & $21)$ & $10^{-13}$ & $(1$ & $589$  & $0$ & $008$ & $0$ & $24)$ & $10^{-13}$ \\
     $3$ & $86$ & $(7$ & $38$  & $0$ & $06$  & $1$ & $2)$  & $10^{-14}$ & $(8$ & $30$   & $0$ & $06$  & $1$ & $3)$  & $10^{-14}$ \\
     $4$ & $86$ & $(3$ & $76$  & $0$ & $05$  & $0$ & $6)$  & $10^{-14}$ & $(4$ & $222$  & $0$ & $04$  & $0$ & $7)$  & $10^{-14}$ \\
     $6$ & $12$ & $(1$ & $868$ & $0$ & $027$ & $0$ & $3)$  & $10^{-14}$ & $(2$ & $098$  & $0$ & $023$ & $0$ & $4)$  & $10^{-14}$ \\
     $7$ & $71$ & $(0$ & $910$ & $0$ & $017$ & $0$ & $15)$ & $10^{-14}$ & $(1$ & $021$  & $0$ & $015$ & $0$ & $20)$ & $10^{-14}$ \\
     $9$ & $70$ & $(4$ & $41$  & $0$ & $10$  & $0$ & $8)$  & $10^{-15}$ & $(4$ & $92$   & $0$ & $10$  & $0$ & $9)$  & $10^{-15}$ \\
    $12$ & $21$ & $(2$ & $13$  & $0$ & $07$  & $0$ & $4)$  & $10^{-15}$ & $(2$ & $38$   & $0$ & $06$  & $0$ & $4)$  & $10^{-15}$ \\
    $15$ & $38$ & $(1$ & $08$  & $0$ & $04$  & $0$ & $18)$ & $10^{-15}$ & $(1$ & $177$  & $0$ & $04$  & $0$ & $20)$ & $10^{-15}$ \\
    $19$ & $36$ & $(5$ & $01$  & $0$ & $26$  & $0$ & $9)$  & $10^{-16}$ & $(5$ & $58$   & $0$ & $23$  & $1$ & $0)$  & $10^{-16}$ \\
    $24$ & $37$ & $(2$ & $45$  & $0$ & $17$  & $0$ & $4)$  & $10^{-16}$ & $(2$ & $66$   & $0$ & $15$  & $0$ & $5)$  & $10^{-16}$ \\
    $30$ & $68$ & $(1$ & $44$  & $0$ & $11$  & $0$ & $22)$ & $10^{-16}$ & $(1$ & $51$   & $0$ & $10$  & $0$ & $23)$ & $10^{-16}$ \\
    $38$ & $62$ & $(7$ & $0$   & $0$ & $7$   & $1$ & $2)$  & $10^{-17}$ & $(7$ & $5$    & $0$ & $7$   & $1$ & $3)$  & $10^{-17}$ \\
    $48$ & $62$ & $(3$ & $6$   & $0$ & $5$   & $0$ & $6)$  & $10^{-17}$ & $(3$ & $72$   & $0$ & $4$   & $0$ & $6)$  & $10^{-17}$ \\
    $61$ & $21$ & $(1$ & $91$  & $0$ & $29$  & $0$ & $29)$ & $10^{-17}$ & $(1$ & $97$   & $0$ & $25$  & $0$ & $3)$  & $10^{-17}$ \\
    $77$ & $06$ & $(1$ & $04$  & $0$ & $18$  & $0$ & $19)$ & $10^{-17}$ & $(1$ & $05$   & $0$ & $17$  & $0$ & $19)$ & $10^{-17}$ \\
    $97$ & $01$ & $(4$ & $6$   & $1$ & $1$   & $1$ & $0)$  & $10^{-18}$ & $(4$ & $7$    & $1$ & $0$   & $1$ & $0)$  & $10^{-18}$ \\
    \bottomrule
  \end{tabular}
  \label{tab:result}
\end{table*}

\begin{table*}%
  \centering%
  \caption{Fit parameters of the cosmic-ray energy spectrum according to function~\eqref{eq:spectrum:hoerandel} for the pure proton and two-component model primary composition assumptions. Systematic errors were derived as described in the text and exclude the systematic error due to the unknown composition, since these are fits of spectra derived under specific composition assumptions.}%
  \subtable[Proton]{%
    \begin{tabular*}{.7\textwidth}{lr@{.}lll}%
      \toprule
      Parameter & \multicolumn{4}{c}{Best fit}                                                                                                  \\
      \midrule
      $I_\mathrm{knee}/10^{-7}\un{m^{-2}\,s^{-1}\,sr^{-1}}$ & $3$  & $8$   & $\pm\; 1.9\mathrm{(stat)}$   & $^{+0.5}_{-1.3}\mathrm{(syst)}$     \\
      $E_\mathrm{knee}/\mathrm{PeV}$                        & $3$  & $2$   & $\pm\; 0.9\mathrm{(stat)}$   & $^{+0.7}_{-0.2}\mathrm{(syst)}$     \\
      $\gamma_1$                                            & $-2$ & $5$   & $\pm\; 0.4\mathrm{(stat)}$   & $^{+0.2}_{-0.7}\mathrm{(syst)}$     \\
      $\gamma_2$                                            & $-3$ & $076$ & $\pm\; 0.019\mathrm{(stat)}$ & $\pm\; 0.15\mathrm{(syst)}$         \\
      $\eps$                                                & \multicolumn{2}{r}{$6$} & $\pm\; 4\mathrm{(stat)}$ &                              \\
      $\chi^2/N_\mathrm{df}$                                & \multicolumn{4}{l}{\phantom{$-$}$15.6/12$}                                        \\
      \bottomrule
    \end{tabular*}%
    \label{tab:spectrum:fitparams_proton}} \\[2ex]%
  \subtable[Two Components]{%
    \begin{tabular*}{.7\textwidth}{lr@{.}lll}
      \toprule
      Parameter & \multicolumn{4}{c}{Best fit}                                                                                           \\
      \midrule
      $I_\mathrm{knee}/10^{-7}\un{m^{-2}\,s^{-1}\,sr^{-1}}$ & $2$  & $38$  & $\pm\; 0.23\mathrm{(stat)}$  & $\pm\; 0.5\mathrm{(syst)}$   \\
      $E_\mathrm{knee}/\mathrm{PeV}$                        & $4$  & $32$  & $\pm\; 0.22\mathrm{(stat)}$  & $\pm\; 0.18\mathrm{(syst)}$  \\
      $\gamma_1$                                            & $-2$ & $759$ & $\pm\; 0.015\mathrm{(stat)}$ & $\pm\; 0.21\mathrm{(syst)}$  \\
      $\gamma_2$                                            & $-3$ & $107$ & $\pm\; 0.016\mathrm{(stat)}$ & $\pm\; 0.3\mathrm{(syst)}$   \\
      $\eps$                                                & \multicolumn{2}{r}{$9$} & $\pm\; 3\mathrm{(stat)}$ &                       \\
      $\chi^2/N_\mathrm{df}$                                & \multicolumn{4}{l}{\phantom{$-$}$19.4/13$}                                 \\
      \bottomrule
    \end{tabular*}%
    \label{tab:spectrum:fitparams_twocomponent}}%
  \label{tab:fitparams}%
\end{table*}

The systematic uncertainty of the knee energy~$E_\mathrm{knee}$ is the systematic error on energy determination at that primary energy as given in Section~\ref{sec:systematics}. 
The systematic error of~$I_\mathrm{knee}$ has been obtained by quadratically adding the systematic error on the flux determination and the systematic energy error transformed into a flux uncertainty based on the local spectral index.
In order to determine systematic errors on~$\gamma_1$ and~$\gamma_2$, the fit was repeated using the systematic errors of the data points as statistical errors.
In case of the proton assumption the systematic uncertainty introduced by the removal of the first two data points from the fit (see above) has been added quadratically to these numbers, which increases the systematic values for this assumption, in particular those of the slope~$\gamma_1$ below the knee and the knee energy.

Above about~$22\un{PeV}$ a possible flattening of the spectrum can be observed independent of primary composition assumption.
This feature is also visible in the measured shower size spectra (see Fig.~\ref{fig:shower_size_spectra}).
In order to test its statistical significance, the spectra were fitted with function~\eqref{eq:spectrum:hoerandel} plus an additional hard break at~$E_\mathrm{break}$ with spectral index~$\gamma_3$.
The goodness of fit improves to~$\chi^2/N_\mathrm{df} = 6.1/10$ for the pure-proton assumption and to~$\chi^2/N_\mathrm{df}=7.1/11$ for the two-component model assumption.
These improvements of the $\chi^2$ correspond to significances of~$2.7$ and~$3.2$ standard deviations respectively, which, however, does not include systematic errors.
The parameters of the flattening are listed in Table~\ref{tab:flattening}.

\begin{table}
  \centering
  \caption{Parameters of the flattening of the spectrum at high energy. The errors given are only statistical.}
  \vspace*{2ex}
  \begin{tabular}{lrlrl}
    \toprule
    Parameter & \multicolumn{2}{c}{Proton} & \multicolumn{2}{c}{Two Components} \\
    \midrule
    $E_\mathrm{break}/\mathrm{PeV}$ & \phantom{$-$}$21$ & $\pm\; 4$    & \phantom{$-$}$23$ & $\pm\; 5$    \\
    $\gamma_3$                      & $-2.82$           & $\pm\; 0.10$ & $-2.87$           & $\pm\; 0.09$ \\
    $\chi^2/N_\mathrm{df}$          & \multicolumn{2}{l}{\phantom{$-$}$6.1/10$} & \multicolumn{2}{l}{\phantom{$-$}$7.1/11$} \\
    \bottomrule
  \end{tabular}
  \label{tab:flattening}
\end{table}

\section{Summary}
We have derived the all-particle cosmic ray energy spectrum in the energy range between~$1\un{PeV}$ and~$100\un{PeV}$ from data taken between June and October 2007 with the 26-station configuration of the IceTop air shower array at South Pole.

Using the air shower simulation package CORSIKA with the high-energy hadronic interaction model SIBYLL2.1 the relation between shower size~$S_{125}$ and primary energy, as well as the detection efficiency and energy resolution were determined.
Three different assumption on the primary mass composition were used as input: pure proton, pure iron and a simple two-component model~\cite{glasstetter}.
Based on these results, shower size spectra obtained in three zenith angle ranges were unfolded with a Bayesian unfolding algorithm to obtain energy spectra.

In case of pure proton and the two-component model, it was found that the spectra obtained in the different zenith angle ranges were in good agreement.
In the pure iron case, on the other hand, a strong disagreement between the three spectra was observed at low energies.
Since one can safely assume that cosmic rays are isotropic in the given energy range, the spectra in all three zenith angle ranges should be the same.
With this assumption, we concluded that pure iron primaries can be excluded below energies of~$25\un{PeV}$.

We showed that the attenuation of air showers with increasing zenith angle bears exploitable information about the chemical composition of cosmic rays.
Nevertheless, the main source of systematic error still remains the primary mass composition.
The systematic error due to the choice of a hadronic interaction model is relatively small in this analysis because most air shower signals are dominated by the electromagnetic component of an air shower, which is relatively well understood.
For the final result, only the spectra obtained from the most vertical zenith angle range, $0\degrees \leq \theta < 30\degrees$, were considered because in this range the dependence on composition and systematic errors are smallest.

In case of the pure-proton assumption the knee in the cosmic-ray energ spectrum was observed at~$3.2\un{PeV}$ with a spectral index of~$-2.5$ below and~$-3.08$ above the knee. 
For the two-component model assumption the knee position was determined at~$4.3\un{PeV}$ with spectral indices of~$-2.76$ below and~$-3.11$ above.
Around an energy of~$22\un{PeV}$ an indication of a flattening of the cosmic ray spectrum to an index of about~$-2.85$ was observed at the~$3\sigma$ level.

Since the completion of IceTop and IceCube in 2011, the array is three times larger than the configuration used in this analysis.
With this larger array, statistics and containment of high-energy showers will be much better, allowing to extend the analysis to higher energies.
The main strength of IceTop, however, is the possibility to measure air showers at the surface in coincidence with high-energy muons penetrating deep enough into the ice to trigger IceCube. 
The ratio between the two measurements is highly sensitive to the mass of the primary particle.

\appendix
\section{Parametrization of the response matrix} \label{app:response_matrix}
In order to mitigate the effects of statistical fluctuations in the unfolding procedure, the response matrices described in Section~\ref{sec:response_matrix} were separated into mean logarithmic shower size $\langle\log S_{125}\rangle$, resolution $\sigma_{\log S}$, and efficiency~$\eps$. There dependences on $x=\log E_p$ were then fitted by empirical functions:
\begin{align}
  \notag\langle\log S_{125}\rangle(x) &= a_0 + x + \\
    \label{eq:s125_function}
    &\quad\ln\left(\frac{\exp(a_1 x) + \exp\bigl(a_2 + a_3 x + a_4 x^2\bigr)}{1 + \exp(a_2)}\right),
\end{align}
\begin{equation} \label{eq:sigma_function}
  \sigma_{\log S}(x) = \frac{b_0\bigl(1 + \exp(b_3b_4)\bigr) + \exp(-b_1)\bigl(\exp(-b_2x)-1\bigr)}{1 + \exp\bigl(-b_3(x - b_4)\bigr)}
\end{equation}
and
\begin{equation}
  \label{eq:efficiency_function}
  \eps(x) = \begin{cases}
    \dfrac{c_0}{1 + \exp\bigl(-c_1(x - c_2) + c_3(x - c_4)^2\bigr)} & x < c_4 \\[2.5ex]
    \dfrac{c_0}{1 + \exp\bigl(-c_1(x - c_2)\bigr)} & x \ge c_4
  \end{cases}.
\end{equation}

\section*{Acknowledgements}
We acknowledge the support from the following agencies: U.S. National Science Foundation-Office of Polar Programs, U.S. National Science Foundation-Physics Division, University of Wisconsin Alumni Research Foundation, the Grid Laboratory Of Wisconsin (GLOW) grid infrastructure at the University of Wisconsin - Madison, the Open Science Grid (OSG) grid infrastructure; U.S. Department of Energy, and National Energy Research Scientific Computing Center, the Louisiana Optical Network Initiative (LONI) grid computing resources; National Science and Engineering Research Council of Canada; Swedish Research Council, Swedish Polar Research Secretariat, Swedish National Infrastructure for Computing (SNIC), and Knut and Alice Wallenberg Foundation, Sweden; German Ministry for Education and Research (BMBF), Deutsche Forschungsgemeinschaft (DFG), Research Department of Plasmas with Complex Interactions (Bochum), Germany; Fund for Scientific Research (FNRS-FWO), FWO Odysseus programme, Flanders Institute to encourage scientific and technological research in industry (IWT), Belgian Federal Science Policy Office (Belspo); University of Oxford, United Kingdom; Marsden Fund, New Zealand; Japan Society for Promotion of Science (JSPS); the Swiss National Science Foundation (SNSF), Switzerland.

\section*{References}
\bibliography{spectrum}
\bibliographystyle{elsarticle-num}

\end{document}